\documentclass[12pt]{article}
\usepackage[utf8]{inputenc}
\usepackage{amsmath,amssymb}
\usepackage{slashed}
\usepackage{graphicx}
\usepackage{cite}
\usepackage{hyperref}
\usepackage{chngcntr}

\usepackage{xcolor}
\usepackage{color}
\newcommand{\eps}{\epsilon}
\newcommand{\Li}{\mathrm{Li}}
\allowdisplaybreaks

\counterwithin*{equation}{section}

\begin{document}

\begin{titlepage}

\vspace*{-2cm}
\begin{flushright}
LU TP 21-03\\
Revised March 2021
\vspace*{2mm}

\end{flushright}

\begin{center}
\vspace*{15mm}

\vspace{1cm}
{\LARGE \bf
The two-loop perturbative correction to the (g-2)$_{\mu}$ HLbL at short distances
} 
\vspace{1.4cm}

\renewcommand{\thefootnote}{\fnsymbol{footnote}}
{Johan Bijnens$^a$, Nils Hermansson-Truedsson$^b$, Laetitia Laub$^{b}$, Antonio Rodr\'iguez-S\'anchez$^{c}$}
\renewcommand{\thefootnote}{\arabic{footnote}}
\setcounter{footnote}{0}

\vspace*{.5cm}
\centerline{$^a${\it Department of Astronomy and Theoretical Physics,
Lund University,}} 
\centerline{\it S\"olvegatan 14A, SE 223-62 Lund, Sweden}
\centerline{${}^b$ \it Albert Einstein Center for Fundamental Physics, Institute for Theoretical Physics, }
\centerline{{\it Universit\"{a}t Bern, Sidlerstrasse 5, CH–3012 Bern, Switzerland}}
\centerline{${}^c$\it Universit\'{e} Paris-Saclay, CNRS/IN2P3, IJCLab, 91405 Orsay,
France }
\vspace*{.2cm}

\end{center}

\vspace*{10mm}
\begin{abstract}\noindent\normalsize
The short-distance behaviour of the hadronic light-by-light (HLbL) contribution to $(g-2)_{\mu}$ has recently been studied by means of an operator product expansion in a background electromagnetic field. The leading term in this expansion has been shown to be given by the massless quark loop, and the non-perturbative corrections are numerically very suppressed. Here, we calculate the perturbative QCD correction to the massless quark loop. The correction is found to be fairly small compared to the quark loop as far as we study energy scales where the perturbative running for the QCD coupling is well-defined, i.e.~for scales $\mu\gtrsim 1\, \mathrm{GeV}$. This should allow to reduce the large systematic uncertainty associated to high-multiplicity hadronic states.
\end{abstract}

\end{titlepage}
\newpage 

\renewcommand{\theequation}{\arabic{section}.\arabic{equation}} 
 
\section{Introduction}\label{sec:intro}

The muon anomalous magnetic moment is one of the most precise measurements in particle physics. The world average \cite{Aoyama:2020ynm} for the anomaly $a_\mu = (g-2)/2$ is
\begin{align}
\label{eq:amuexp}
a_\mu^{\textrm{exp}} =      116~592~089(54)(33)\times 10^{-11}\,.
\end{align}
The experimental accuracy is expected to improve with the now running experiment at Fermilab \cite{Grange:2015fou} and the planned experiment at J-PARC \cite{Abe:2019thb}. The Standard Model prediction \cite{Aoyama:2020ynm} is
\begin{align}
\label{eq:amuSM}
    a_\mu^{\textrm{SM}} = 116~591~810(43) \times 10^{-11}\,.
\end{align}
The difference between this and the experimental value from Brookhaven National Laboratory~\cite{Bennett:2006fi} is
\begin{align}
    \Delta a_\mu \equiv a_\mu^{\textrm{exp}}-a_\mu^{\textrm{SM}} = 279(76)\times 10^{-11} \,,
\end{align}
or a $3.7\sigma$ discrepancy. In light of this discrepancy and the expected improved experimental accuracy it is important that the theoretical accuracy is checked as much as possible.
The QED \cite{Aoyama:2012wk,Aoyama:2019ryr} and the electroweak contribution \cite{Czarnecki:2002nt,Gnendiger:2013pva} are precise enough for the foreseeable future. The error is dominated by the hadronic contributions, the hadronic vacuum polarization
\cite{Davier:2017zfy,Keshavarzi:2018mgv,Colangelo:2018mtw,Hoferichter:2019gzf,Davier:2019can,Keshavarzi:2019abf,Kurz:2014wya} is at present the largest theory uncertainty but is steadily being improved. The remaining part, the hadronic light-by-light (HLbL) contribution is at present \cite{Aoyama:2020ynm}
\begin{align}
\label{eq:amuHLbL}
    a_\mu^{\textrm{HLbL}} = 92(18)\times 10^{-11}.
\end{align}
This number contains the next-to-leading (NLO) HLbL contribution \cite{Colangelo:2014qya} and the average of the lattice \cite{Blum:2019ugy} and phenomenological evaluation of the lowest-order HLbL. In the remainder we will use HLbL as a synonym for the LO part only, this contribution is depicted in Fig.~\ref{fig:hlbl}. The number in~(\ref{eq:amuHLbL}) is in good agreement with the older estimates \cite{Bijnens:1995cc,Bijnens:1995xf,Hayakawa:1997rq,Bijnens:2001cq,Hayakawa:2001bb} and the more recent Glasgow consensus \cite{Prades:2009tw} but with a smaller and much better understood error.

The phenomenological estimate of the HLbL \cite{Aoyama:2020ynm},
\begin{align}
    a_\mu^{\textrm{HLbL-phen}} = 92(19)\times 10^{-11},
\end{align}
uses the methods of Ref.~\cite{Colangelo:2015ama} to separate different contributions. The pole contributions from $\pi^0,\eta,\eta'$ 
\cite{Masjuan:2017tvw,Hoferichter:2018kwz,Gerardin:2019vio}
as well as the two-pion box and rescattering and two-kaon box contribution \cite{Colangelo:2017fiz} are well-understood and together give
\begin{align}
    a_\mu^{\textrm{HLbL-1}} = 69.4(4.1)\times 10^{-11}\,.
\end{align}
The main uncertainty comes from the intermediate and short-distance domain. Heavier intermediate states have been considered in Refs.~\cite{Pauk:2014rta,Danilkin:2016hnh,Jegerlehner:2017gek,Knecht:2018sci,Eichmann:2019bqf,Roig:2019reh}. The heavy-quark contribution from charm is sufficiently well estimated from the quark loop and estimates of non-perturbative contributions and that of the bottom and top quarks are negligible \cite{Aoyama:2020ynm,Kuhn:2003pu,Colangelo:2019uex,Colangelo:2019lpu}.
The light-quark contribution can be estimated using the quark loop and/or higher resonance exchanges and leads to \cite{Aoyama:2020ynm}
\begin{align}
    a_\mu^{\textrm{HLbL-SD1}} = 20(19)\times 10^{-11}.
\end{align}
The large error is due to the large uncertainty of which resonances to include and that their couplings to two off-shell photons are badly known~\cite{Aoyama:2020ynm}. In addition one needs to make sure that there is a proper matching with the short-distance QCD constraints.

Some short-distance constraints are used in determining the form-factors needed in the contributions from hadrons directly, see e.g.~Ref.~\cite{Knecht:2001qf}. Here we discuss instead the short-distance constraints on the hadronic function defined in~(\ref{eq:hlbltensor}) and depicted as the shaded blob in Fig.~\ref{fig:hlbl}. First attempts at matching the short-distance were using the quark loop and matching it on a long-distance contribution from the extended Nambu-Jona-Lasinio model~\cite{Bijnens:1995xf}. The quark loop itself has a long history of being used in this context, see e.g.~Ref.~\cite{Kinoshita:1984it,Goecke:2010if,Boughezal:2011vw,Greynat:2012ww,Masjuan:2012qn,Dorokhov:2015psa}.
The first proper short-distance constraint was derived in Ref.~\cite{Melnikov:2003xd}.
It is valid in the regime where two of the internal photons have a virtuality much larger than the third one. Recent work in the latter regime includes Refs.~\cite{Colangelo:2019lpu,Colangelo:2019uex,Melnikov:2019xkq,Leutgeb:2019gbz,Cappiello:2019hwh,Knecht:2020xyr,Masjuan:2020jsf,Ludtke:2020moa,Hoferichter:2020lap}. 

This paper is concerned with the limit where all virtualities of the internal photon lines in Fig.~\ref{fig:hlbl} are large. The underlying problem here is that the external photon, corresponding to the magnetic field, has zero momentum, i.e.~$q_4\to0$ in Fig.~\ref{fig:hlbl}. The usual operator product expansion (OPE) in vacuum~\cite{Shifman:1978bx} corresponds to all four photon virtualities large and diverges when setting $q_4\to0$. The solution was found in Ref.~\cite{Bijnens:2019ghy}. One needs to use an alternative OPE in a background magnetic field as was done for the QCD sum rule calculations of nucleon magnetic moments \cite{Balitsky:1983xk,Ioffe:1983ju}. This method was earlier used in the context of the electroweak contribution to $a_\mu$ \cite{Czarnecki:2002nt}. The first order term in this expansion corresponds to the massless quark loop \cite{Bijnens:2019ghy}, the next order is suppressed by quark masses and the small value of the magnetic susceptibility \cite{Bijnens:2019ghy,Bijnens:2020xnl}. For the non-perturbative part of this OPE the contribution suppressed by up to four powers of large momenta compared to the leading term have been evaluated in Ref.~\cite{Bijnens:2020xnl}. There are a number of subtleties involved and large number of expectation values in a magnetic field needed to be evaluated. The conclusion from \cite{Bijnens:2019ghy,Bijnens:2020xnl} is that the contribution from these higher orders in the non-perturbative part are small. The remaining uncertainty from this regime is the perturbative correction from gluon exchange to the massless quark loop. This paper performs that calculation. The putting together of this work with the other short-distance constraint~\cite{Melnikov:2003xd} and the parts calculated using hadronic methods is deferred to future work.

In Sec.~\ref{sec:HLbLtensor} we recall the main definitions needed for the calculation of the HLbL part of $a_\mu$. We define here a set of intermediate quantities, the $\tilde\Pi_i$ that are both ultraviolet and infrared finite. From these we then determine the quantities $\hat\Pi_i$ that are needed to calculate $a_\mu$. The main procedure of the calculation is described in Sec.~\ref{sec:2loop}. Sec.~\ref{sec:numerics} gives the numerical results and discusses implications. We reiterate our main results in Sec.~\ref{sec:concl}. A number of technical issues are relegated to the the appendices. The final result is too large to include in the manuscript but is included as supplementary material~\cite{supplementary}.

\section{The HLbL tensor and $a_{\mu}^{\textrm{HLbL}}$}\label{sec:HLbLtensor}

\begin{figure}[tb]\centering
\includegraphics[width=0.25\textwidth]{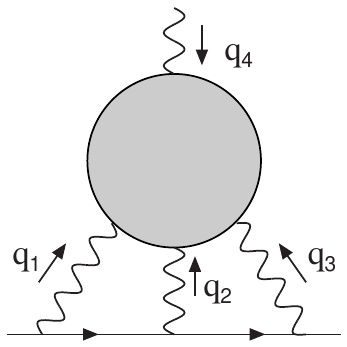}
\caption{\label{fig:hlbl}The HLbL contribution to the $(g-2)_{\mu}$.}
\end{figure}

The HLbL tensor $\Pi^{\mu_{1}\mu_{2}\mu_{3}\mu_{4} }$ is a 4-point correlation function of electromagnetic currents $J^{\mu}(x) = \bar{q}(x)\, Q_{q}\gamma ^{\mu}q(x)$,
where the quark fields are collected in $q=(u,d,s)$ and the corresponding charge matrix is $Q_{q}=\textrm{diag}(e_q) =\textrm{diag}(2/3,-1/3,-1/3)$. The correlator in question is defined via 
\begin{equation}\label{eq:hlbltensor}
\Pi^{\mu_{1}\mu_{2}\mu_{3}\mu_{4} } 
=
-i\int \frac{d^{4}q_{4}}{(2\pi)^{4}}\left(\prod_{i=1}^{4}\int d^{4}x_{i}\, e^{-i q_{i} x_{i}}\right)  \langle 0 | T\left(\prod_{j=1}^{4}J^{\mu_{j}}(x_{j})\right)|0\rangle \, ,
\end{equation}
where the $q_{i}$ are the momenta of the external photon legs. This definition is slightly unconventional but allows to exploit more of the symmetries, as remarked in Ref.~\cite{Bijnens:2020xnl}. The contribution from the HLbL tensor to the $(g-2)_{\mu}$ is depicted in Fig.~\ref{fig:hlbl}. It involves a loop integration over $q_{1}$, $q_{2}$ and $q_{3}$, whereas the fourth leg is in the static limit, i.e.~$q_{4}\rightarrow 0$.

The HLbL tensor satisfies the Ward identities
$q_{i,\, \mu_{i}} \, \Pi^{\mu_{1}\mu_{2}\mu_{3}\mu_{4}}=0$  for $i = 1,2,3,4 \, ,$
which implies~\cite{Aldins:1970id}
\begin{equation}\label{eq:wardcons}
\Pi^{\mu_{1}\mu_{2}\mu_{3}\mu_{4}}=-q_{4,\, \nu_{4}}\frac{\partial \Pi^{\mu_{1}\mu_{2}\mu_{3}\nu_{4}}}{\partial q_{4,\, \mu_{4}}} \, .
\end{equation}
The whole information about the HLbL is then contained in its derivative. In fact, in the $(g-2)_{\mu}$ kinematics,
\begin{equation}
\lim_{q_4\to 0} \frac{\partial \Pi^{\mu_{1} \mu_{2} \mu_{3} \nu_{4}}}{\partial q_{4}^{\mu_{4}}} \, ,
\end{equation}
there are only $19$ independent Lorentz structures, which can be found by applying $19$ independent projectors $P^{\tilde{\Pi}_{i}}_{\mu_1 \mu_2 \mu_3 \mu_4 \nu_4}$, this according to
\begin{equation}\label{eq:pitildes}
\tilde{\Pi}_{i}=P^{\tilde{\Pi}_{i}}_{\mu_1 \mu_2 \mu_3 \mu_4 \nu_4}\lim_{q_4\to 0} \frac{\partial \Pi^{\mu_{1} \mu_{2} \mu_{3} \nu_{4}}}{\partial q_{4}^{\mu_{4}}} \, .
\end{equation}
A possible set of projectors is
\begin{align}\label{eq:firstproj}
P^{\tilde{\Pi}_{1}}_{\mu_1 \mu_2 \mu_3 \mu_4 \nu_4}&=g_{\mu_1\mu_2}\,g_{\mu_3\nu_4}\;q_{1,\mu_4}\, ,\\
P^{\tilde{\Pi}_{2}}_{\mu_1 \mu_2 \mu_3 \mu_4 \nu_4}&=g_{\mu_2\mu_3}\,g_{\mu_1\nu_4}\;q_{2,\mu_4}\, ,\\
P^{\tilde{\Pi}_{3}}_{\mu_1 \mu_2 \mu_3 \mu_4 \nu_4}&=g_{\mu_3\mu_1}\,g_{\mu_2\nu_4}\;q_{3,\mu_4}\, ,\\
P^{\tilde{\Pi}_{4}}_{\mu_1 \mu_2 \mu_3 \mu_4 \nu_4}&=g_{\mu_2\mu_1}\,g_{\mu_3\nu_4}\;q_{2,\mu_4}\, ,\\
P^{\tilde{\Pi}_{5}}_{\mu_1 \mu_2 \mu_3 \mu_4 \nu_4}&=g_{\mu_3\mu_2}\,g_{\mu_1\nu_4}\;q_{3,\mu_4}\, ,\\
P^{\tilde{\Pi}_{6}}_{\mu_1 \mu_2 \mu_3 \mu_4 \nu_4}&=g_{\mu_1\mu_3}\,g_{\mu_2\nu_4}\;q_{1,\mu_4}\, ,\\[0.5cm]
P^{\tilde{\Pi}_{7}}_{\mu_1 \mu_2 \mu_3 \mu_4 \nu_4}&=g_{\mu_1\nu_4}\,g_{\mu_2\mu_4}\;q_{2,\mu_3}\, ,\\
P^{\tilde{\Pi}_{8}}_{\mu_1 \mu_2 \mu_3 \mu_4 \nu_4}&=g_{\mu_2\nu_4}\,g_{\mu_3\mu_4}\;q_{3,\mu_1}\, ,\\
P^{\tilde{\Pi}_{9}}_{\mu_1 \mu_2 \mu_3 \mu_4 \nu_4}&=g_{\mu_3\nu_4}\,g_{\mu_1\mu_4}\;q_{1,\mu_2}\,   ,\\[0.5cm]
P^{\tilde{\Pi}_{10}}_{\mu_1 \mu_2 \mu_3 \mu_4 \nu_4}&=g_{\mu_1\mu_2}\,q_{1,\mu_3}q_{1,\nu_4}\;q_{2,\mu_4}\, ,\\
P^{\tilde{\Pi}_{11}}_{\mu_1 \mu_2 \mu_3 \mu_4 \nu_4}&=g_{\mu_2\mu_3}\,q_{2,\mu_1}q_{2,\nu_4}\;q_{3,\mu_4}\, ,\\
P^{\tilde{\Pi}_{12}}_{\mu_1 \mu_2 \mu_3 \mu_4 \nu_4}&=g_{\mu_3\mu_1}\,q_{3,\mu_2}q_{3,\nu_4}\;q_{1,\mu_4}\,   ,\\[0.5cm]
P^{\tilde{\Pi}_{13}}_{\mu_1 \mu_2 \mu_3 \mu_4 \nu_4}&=g_{\mu_1\nu_4}\,q_{1,\mu_2}q_{2,\mu_3}\;q_{3,\mu_4}\, ,\\
P^{\tilde{\Pi}_{14}}_{\mu_1 \mu_2 \mu_3 \mu_4 \nu_4}&=g_{\mu_2\nu_4}\,q_{2,\mu_3}q_{3,\mu_1}\;q_{1,\mu_4}\, ,\\
P^{\tilde{\Pi}_{15}}_{\mu_1 \mu_2 \mu_3 \mu_4 \nu_4}&=g_{\mu_3\nu_4}\,q_{3,\mu_1}q_{1,\mu_2}\;q_{2,\mu_4}\,  ,\\
P^{\tilde{\Pi}_{16}}_{\mu_1 \mu_2 \mu_3 \mu_4 \nu_4}&=g_{\mu_2\nu_4}\,q_{2,\mu_1}q_{1,\mu_3}\;q_{3,\mu_4}\, ,\\
P^{\tilde{\Pi}_{17}}_{\mu_1 \mu_2 \mu_3 \mu_4 \nu_4}&=g_{\mu_3\nu_4}\,q_{3,\mu_2}q_{2,\mu_1}\;q_{1,\mu_4}\, ,\\
P^{\tilde{\Pi}_{18}}_{\mu_1 \mu_2 \mu_3 \mu_4 \nu_4}&=g_{\mu_1\nu_4}\,q_{1,\mu_3}q_{3,\mu_2}\;q_{2,\mu_4}\,  ,\\[0.5cm]
P^{\tilde{\Pi}_{19}}_{\mu_1 \mu_2 \mu_3 \mu_4 \nu_4}&=q_{3,\mu_1}q_{1,\mu_2}\,q_{2,\mu_3}q_{1,\nu_4}\;q_{2,\mu_4}\, ,
\end{align}
which has been built in such a way that, combined with the crossing symmetries of the HLbL tensor, the $\tilde{\Pi}$ satisfy the following crossing symmetries
\begin{align}\nonumber
&\tilde{\Pi}_{1}=C_{12}[\tilde{\Pi}_{4}], \tilde{\Pi}_{2}=C_{12}[\tilde{\Pi}_{6}],
\tilde{\Pi}_{3}=C_{12}[\tilde{\Pi}_{5}], \tilde{\Pi}_{7}=C_{12}[\tilde{\Pi}_{7}],
\tilde{\Pi}_{8}=C_{12}[\tilde{\Pi}_{9}],
\tilde{\Pi}_{10}=C_{12}[\tilde{\Pi}_{10}],\\ \nonumber
&\tilde{\Pi}_{11}=C_{12}[\tilde{\Pi}_{12}], 
\tilde{\Pi}_{13}=C_{12}[\tilde{\Pi}_{16}],
\tilde{\Pi}_{14}=C_{12}[\tilde{\Pi}_{18}],
\tilde{\Pi}_{15}=C_{12}[\tilde{\Pi}_{17}],
\tilde{\Pi}_{19}=C_{12}[\tilde{\Pi}_{19}],\\ \nonumber
&\tilde{\Pi}_{1}=C_{13}[\tilde{\Pi}_{5}],
\tilde{\Pi}_{2}=C_{13}[\tilde{\Pi}_{4}], 
\tilde{\Pi}_{3}=C_{13}[\tilde{\Pi}_{6}],
\tilde{\Pi}_{7}=C_{13}[\tilde{\Pi}_{8}],
\tilde{\Pi}_{9}=C_{13}[\tilde{\Pi}_{9}],
\tilde{\Pi}_{10}=C_{13}[\tilde{\Pi}_{11}],\\ 
&\tilde{\Pi}_{12}=C_{13}[\tilde{\Pi}_{12}],
\tilde{\Pi}_{13}=C_{13}[\tilde{\Pi}_{17}],
\tilde{\Pi}_{14}=C_{13}[\tilde{\Pi}_{16}],
\tilde{\Pi}_{15}=C_{13}[\tilde{\Pi}_{18}],
\tilde{\Pi}_{19}=C_{13}[\tilde{\Pi}_{19}].\label{eq:crosstilde}
\end{align}
The operator $C_{ij}$ interchanges two momenta $q_i$ and $q_j$. 
Notice how, from the knowledge of five of them, for example $\tilde{\Pi}_{1,7,10,13,19}$, one can easily infer the rest from these crossing symmetries. These $\tilde\Pi_i$ are well-defined and
are both ultraviolet and infrared finite to the order in $\alpha$ we are working. For our calculation we use two different sets of $\tilde\Pi_i$, related by gauge invariance, which thus provides a cross-check of our results.

An OPE is only valid for large Euclidean momenta~\cite{Shifman:1978bx}. As a consequence, it cannot be directly applied to the tensor in~(\ref{eq:hlbltensor}) for the $(g-2)_{\mu}$ kinematics, since by definition the external photon is soft, $q_{4}\rightarrow 0$, even though the other Euclidean momenta are large, $-q_{i}^2 \equiv Q_{i}^2\gg\Lambda_{\mathrm{QCD}}^2$~\cite{Bijnens:2019ghy,Bijnens:2020xnl}. However, precisely the same fact allows one to connect the tensor in~(\ref{eq:hlbltensor}) to the OPE of the tensor operator with the background photon field
\begin{align}
\label{eq:backdyson1}
&
\Pi ^{\mu_{1} \mu_{2} \mu_{3} }(q_{1},q_{2})
=
-\frac{1}{e}\int\frac{d^4 q_{3}}{(2\pi)^4}  \left(\prod_{i=1}^{3}\int d^{4}x_{i}\, e^{-i q_{i} x_{i}}\right)  \langle 0 | T\left(\prod_{j=1}^{3}J^{\mu_{j}}(x_{j})\right) | \gamma(q_4) \rangle \, .
\end{align}
The OPE in question holds for large photon virtualities $Q_{1}^2\sim Q_{2}^{2} \sim Q_{3}^{2}\gg\Lambda_{\mathrm{QCD}}^2$.
In this expansion, any local operator with the same quantum numbers as $F_{\mu\nu}$, including $F_{\mu\nu}$ itself, can absorb the remaining soft static photon and, as a consequence, give a contribution \cite{Czarnecki:2002nt,Bijnens:2019ghy,Bijnens:2020xnl}. Higher-dimensional operators are suppressed by extra powers of $\left(\frac{\Lambda_{\mathrm{QCD}}}{Q_{i}}\right)^{d}$, providing a hierarchy of contributions with a systematic counting. A very detailed study of this OPE can be found in Ref.~\cite{Bijnens:2020xnl}, where the different power corrections were computed and found to be small compared to the leading contribution\footnote{Obviously, the results of this expansion cannot be applied to the whole integral domain of~(\ref{eq:amuint}), but it can be used, apart from matching resonance models, for directly evaluating the significant contributions coming from the regions were the OPE is valid.}. The leading term comes from the $F_{\mu\nu}$ operator itself and is given by the massless quark loop at order $\alpha _{s}^0$, and the leading mass effects are very small. In fact, this quark loop corresponds to the zero momentum limit of the derivative of the naive massless perturbative QCD tensor of~(\ref{eq:hlbltensor}), i.e.
\begin{equation}
\lim_{q_4\to 0} \frac{\partial \Pi^{\mu_{1} \mu_{2} \mu_{3} \nu_{4}}_{\mathrm{pert}}}{\partial q_{4}^{\mu_{4}}}    \, .
\end{equation}
In this work, we compute the leading $\alpha_{s}$ correction to the direct $F_{\mu\nu}$ contribution in the OPE of~(\ref{eq:backdyson1}). This corresponds to a two-loop massless QCD calculation with three external legs off-shell.

Before discussing the gluonic correction to the quark loop in the next section, we first write down the quark loop result for the $\tilde{\Pi}_{i}$ basis. As will be remarked upon below, this basis was not used in Ref.~\cite{Bijnens:2020xnl}.
The whole solution for the quark loop can be simply written as
\begin{align}\nonumber
\label{eq:pitilde1quark}
\frac{\pi^2\tilde{\Pi}_{1}}{N_{c}e_{q}^4}
&=C_{123}(0)\left( Q_{3}^{2} - Q_{1}^{2}\right)- 1 + Q_{2}^{-2}\,Q_{3}^{2} + 2\,Q_{2}^{2}\,Q_{3}^{-2} - Q_{1}^{2}\,Q_{2}^{-2} - 2
         \,Q_{1}^{2}\,Q_{3}^{-2} \\ \nonumber
&+ \log \frac{Q_{2}^2}{Q_{3}^2}\, \left( 1 - Q_{2}^{2}\,Q_{3}^{-2} - Q_{1}^{2}\,Q_{3}^{-2} \right)\\ 
&+ \log \frac{Q_{1}^2}{Q_{3}^2}\, \left(  - \frac{5}{2} + \frac{1}{2}\,Q_{2}^{-2}\,Q_{3}^{2} + Q_{2}^{2}\,
Q_{3}^{-2} + \frac{1}{2}\,Q_{1}^{2}\,Q_{2}^{-2} + Q_{1}^{2}\,Q_{3}^{-2} \right) \, ,\\ 
\frac{\pi^2\tilde{\Pi}_{7}}{N_{c}e_{q}^4}
&=C_{123}(0)\left( Q_{3}^{2} - Q_{2}^{2} - Q_{1}^{2} \right) \nonumber \\
&+ \log \frac{Q_{2}^2}{Q_{3}^2}\, \left(  - 1 - Q_{2}^{2}\,Q_{3}^{-2} + Q_{1}^{2}\,Q_{3}^{-2} \right) 
+ \log \frac{Q_{1}^2}{Q_{3}^2}\, \left(  - 1 + Q_{2}^{2}\,Q_{3}^{-2} - Q_{1}^{2}\,Q_{3}^{-2} \right) \, ,\\ \nonumber
\frac{\pi^2\tilde{\Pi}_{10}}{N_{c}e_{q}^4}
&=C_{123}(0)\left( \frac{1}{2}\,Q_{3}^4 - \frac{1}{2}\,Q_{2}^{2}\,Q_{3}^{2} - \frac{1}{2}\,Q_{1}^{2}\,Q_{3}^{2} + Q_{1}^{2}\,Q_{2}^{2} \right) 
\\ \nonumber
&- \frac{1}{2}\,Q_{3}^{2} + Q_{2}^{2} - \frac{1}{2}\,Q_{2}^4\,Q_{3}^{-2} + Q_{1}^{2} + Q_{1}^{2}\,
         Q_{2}^{2}\,Q_{3}^{-2} - \frac{1}{2}\,Q_{1}^4\,Q_{3}^{-2}  \\ \nonumber
&+ \log \frac{Q_{2}^2}{Q_{3}^2}\, \left(  - \frac{3}{4}\,Q_{3}^{2} - \frac{1}{2}\,Q_{2}^{2} + \frac{1}{4}\,Q_{2}^4\,
         Q_{3}^{-2} + Q_{1}^{2} - \frac{1}{4}\,Q_{1}^4\,Q_{3}^{-2} \right)
         \\ 
&+ \log \frac{Q_{1}^2}{Q_{3}^2}\, \left(  - \frac{3}{4}\,Q_{3}^{2} + Q_{2}^{2} - \frac{1}{4}\,Q_{2}^4\,
         Q_{3}^{-2} - \frac{1}{2}\,Q_{1}^{2} + \frac{1}{4}\,Q_{1}^4\,Q_{3}^{-2} \right) \, ,\\ \nonumber
\frac{\pi^2\tilde{\Pi}_{13}}{N_{c}e_{q}^4}
&=C_{123}(0)\left( \frac{1}{2}\,Q_{3}^4 - \frac{1}{2}\,Q_{2}^4 - Q_{1}^{2}\,Q_{3}^{2} + \frac{1}{2}\,Q_{1}^{2}\,
         Q_{2}^{2} \right)\\ \nonumber
&+  \frac{1}{4}\,Q_{2}^{-2}\,Q_{3}^4 - \frac{1}{2}\,Q_{3}^{2} + \frac{1}{4}\,Q_{2}^{2} - \frac{1}{2}\,Q_{1}^{2}\,
         Q_{2}^{-2}\,Q_{3}^{2} - \frac{1}{2}\,Q_{1}^{2} + \frac{1}{4}\,Q_{1}^4\,Q_{2}^{-2} \\
&+ \log \frac{Q_{2}^2}{Q_{3}^2} \, \left(  - \frac{1}{4}\,Q_{3}^{2} - \frac{7}{4}\,Q_{2}^{2} + \frac{3}{4}\,Q_{1}^{2} \right) \nonumber \\
&+ \log \frac{Q_{1}^2}{Q_{3}^2} \, \left(  - Q_{3}^{2} + Q_{2}^{2} + \frac{1}{4}\,Q_{1}^{2}\,Q_{2}^{-2}\,
         Q_{3}^{2} + \frac{1}{4}\,Q_{1}^{2} - \frac{1}{4}\,Q_{1}^4\,Q_{2}^{-2} \right) \, ,\\ \nonumber
\frac{\pi^2\tilde{\Pi}_{19}}{N_{c}e_{q}^4}&=C_{123}(0)\Big(  - \frac{1}{4}\,Q_{3}^6 + \frac{1}{4}\,Q_{2}^{2}\,Q_{3}^4 + \frac{1}{4}\,Q_{2}^4\,Q_{3}^{2}
- \frac{1}{4}\,Q_{2}^6 + \frac{1}{4}\,Q_{1}^{2}\,Q_{3}^4 - \frac{3}{2}\,Q_{1}^{2}\,Q_{2}^{2}\,Q_{3}^{2} + \frac{1}{4}\,Q_{1}^{2}\,
Q_{2}^4 \\& \nonumber + \frac{1}{4}\,Q_{1}^4\,Q_{3}^{2} + \frac{1}{4}\,Q_{1}^4\,Q_{2}^{2} - \frac{1}{4}\,Q_{1}^6 \Big)\\ \nonumber
&+   \frac{1}{2}\,Q_{3}^4 - Q_{2}^{2}\,Q_{3}^{2} + \frac{1}{2}\,Q_{2}^4 - Q_{1}^{2}\,Q_{3}^{2} - Q_{1}^{2}\,
         Q_{2}^{2} + \frac{1}{2}\,Q_{1}^4  \\ \nonumber
&+ \log \frac{Q_{2}^2}{Q_{3}^2} \, \left( \frac{1}{2}\,Q_{3}^4 + \frac{1}{2}\,Q_{2}^{2}\,Q_{3}^{2} - Q_{2}^4
          - Q_{1}^{2}\,Q_{3}^{2} + \frac{1}{2}\,Q_{1}^{2}\,Q_{2}^{2} + \frac{1}{2}\,Q_{1}^4 \right)\\ 
&+ \log \frac{Q_{1}^2}{Q_{3}^2} \, \left( \frac{1}{2}\,Q_{3}^4 - Q_{2}^{2}\,Q_{3}^{2} + \frac{1}{2}\,Q_{2}^4
          + \frac{1}{2}\,Q_{1}^{2}\,Q_{3}^{2} + \frac{1}{2}\,Q_{1}^{2}\,Q_{2}^{2} - Q_{1}^4 \right) \, .
\label{eq:pitilde19quark}
\end{align}
Here $N_{c}$ is the number of colours and $C_{123}(0)$ is a loop integral function that is defined in App.~\ref{app:mi}.

For the $(g-2)_{\mu}$ integration, it is convenient using the generic results of Refs.~\cite{Colangelo:2015ama,Colangelo:2017fiz,Bijnens:2020xnl}. Following them, the HLbL tensor can be expanded in a basis of 54 scalar functions $\Pi _{i}$ weighted with Lorentz structures $T_{i}^{\mu _{1} \mu _{2} \mu _{3} \mu _ {4}}$,
\begin{equation}
\Pi ^{\mu_{1} \mu_{2} \mu_{3} \mu_{4}} = \sum _{i=1}^{54}\, T_{i}^{\mu_{1} \mu_{2} \mu_{3} \mu_{4}} \Pi _{i} \,.
\end{equation}
Using
\begin{equation}\label{eq:derq4btt}
\lim_{q_4\to 0} \frac{\partial \Pi^{\mu_{1} \mu_{2} \mu_{3} \nu_{4}}}{\partial q_{4}^{\mu_{4}}}=\lim_{q_4\to 0} \sum_{i=1}^{54}\frac{\partial T_i^{\mu_{1} \mu_{2} \mu_{3} \nu_{4}}}{\partial q_{4}^{\mu_{4}}}\,\Pi_{i} \, ,
\end{equation}
the $19$ $\tilde{\Pi}_{i}$ defined in~(\ref{eq:pitildes}) can be identified with the static $q_{4}\rightarrow 0$ limit of certain linear combinations of the $\Pi_{i}$. Denoting these linear combinations $\hat{\Pi}_{i}$ it can further be shown that for $a_{\mu}^{\textrm{HLbL}}$ only six $\hat{\Pi}_{i}$ contribute, namely  $\hat{\Pi}_{1,4,7,17,39,54}$.\footnote{Using the set of projectors defined in Ref.~\cite{Bijnens:2020xnl}, the identification of these $\hat{\Pi}_{i}$ as combinations of the $\tilde{\Pi}_{i}$ is straightforward.} In particular, the $a_{\mu}^{\textrm{HLbL}}$ may be written~\cite{Colangelo:2015ama,Colangelo:2017fiz}
\begin{align}\label{eq:amuint}
a_{\mu}^{\textrm{HLbL}} 
= 
\frac{2\alpha ^{3}}{3\pi ^{2}} 
& \int _{0}^{\infty} dQ_{1}\int_{0}^{\infty} dQ_{2} \int _{-1}^{1}d\tau \, \sqrt{1-\tau ^{2}}\, Q_{1}^{3}Q_{2}^{3}
\sum _{i=1}^{12} T_{i}(Q_{1},Q_{2},\tau)\, \overline{\Pi}_{i}(Q_{1},Q_{2},\tau)\, .
\end{align}
The integration variable $\tau$ is defined via $Q_3^2=Q_1^2+Q_2^2+2\tau \, Q_1Q_2$, the $T_{i}(Q_{1},Q_{2},\tau)$ are functions and the $\overline{\Pi}_{i}$ are functions of the six $\hat{\Pi}_{i}$. The latter set of functions is related to the $\hat{\Pi}_{i}$ through
\begin{align}\label{eq:pibarfcns}
    & \overline{\Pi}_{1} = \hat{\Pi}_{1} \, , \; \overline{\Pi}_{2} = C_{23}\left[  \hat{\Pi}_{1}\right] \, , \; \overline{\Pi}_{3} = \hat{\Pi}_{4} \, , \; \overline{\Pi}_{4} = C_{23}\left[\hat{\Pi}_{4}\right]\, , 
    \nonumber \\
    & \overline{\Pi}_{5} = \hat{\Pi}_{7} \, , \; \overline{\Pi}_{6} = C_{12}\left[ C_{13}\left[  \hat{\Pi}_{7}\right] \right] \, , \; \overline{\Pi}_{7} = C_{23}\left[\hat{\Pi}_{7}\right] \, , \; 
    \nonumber  \\
    & \overline{\Pi}_{8} = C_{13}\left[\hat{\Pi}_{17}\right]\, , \; 
     \overline{\Pi}_{9} = \hat{\Pi}_{17} \, , \; \overline{\Pi}_{10} = \hat{\Pi}_{39} \, , \; 
     \nonumber  \\
    & \overline{\Pi}_{11} = -C_{23}\left[ \hat{\Pi}_{54}\right]  \, , \; \overline{\Pi}_{12} = \hat{\Pi}_{54}\, .
\end{align}

In summary, knowledge of $\hat{\Pi}_{1,4,7,17,39,54}$ is enough to determine $a_{\mu}^{\mathrm{HLbL}}$ from (\ref{eq:amuint}). The $\hat\Pi_i$ can be obtained from the derivative of the HLbL tensor in the static limit with the projectors given in Ref.~\cite{Bijnens:2020xnl}. There we defined
\begin{align}
    \hat\Pi_i = P_{\hat\Pi_i \mu_1\mu_2\mu_3\mu_4\nu_4}\lim_{q_4\to0}
    \frac{\partial\Pi^{\mu_1\mu_2\mu_3\nu_4}}{\partial q_{4,\mu_4}}\,.
\end{align}
with the projectors $ P_{\hat\Pi_i \mu_1\mu_2\mu_3\mu_4\nu_4}$ given in App.~A of \cite{Bijnens:2020xnl}. Using the definitions of the $\tilde\Pi$ in (\ref{eq:pitildes}) the relation between the $\hat\Pi$ and the $\tilde\Pi$ follows immediately.
We have checked that this procedure reproduces the massless
quark loop results as given in Ref.~\cite{Bijnens:2020xnl}. The $\tilde{\Pi}_{i}$ representation of the massless quark loop was given in~(\ref{eq:pitilde1quark})--(\ref{eq:pitilde19quark}), and it can be noted that it is much simpler than the expressions for the $\hat\Pi_i$ in Ref.~\cite{Bijnens:2020xnl}.

\section{The two-loop perturbative correction}\label{sec:2loop}
In this section we present the calculation of the two-loop contribution. For the analytic calculation we use {\sc FORM}~\cite{Vermaseren:2000nd}. The master integral reduction is done by means of {Kira}~\cite{Maierhoefer:2017hyi}, which employs a Laporta algorithm to reach a minimal set of master integrals. Explicit analytic expressions of the master integrals can be found in the literature.

\begin{figure}[tb]\centering
\includegraphics[width=0.65\textwidth]{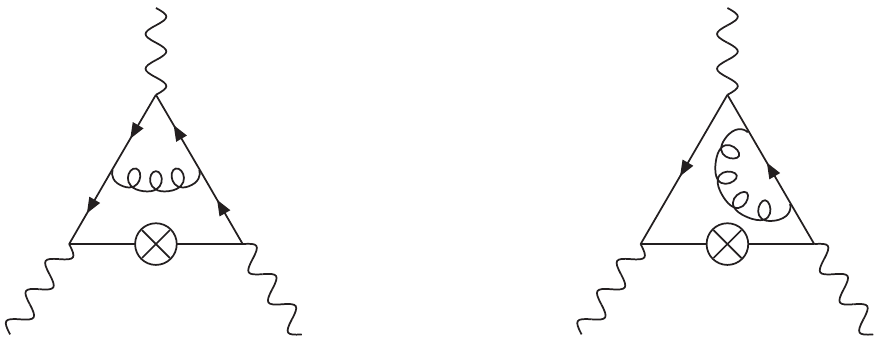}
\caption{\label{fig:2looptop} Two examples of the two-loop perturbative topologies. The external static photon has been indicated by a crossed vertex. }
\end{figure}

\begin{figure}[t]
  \centering
  \includegraphics{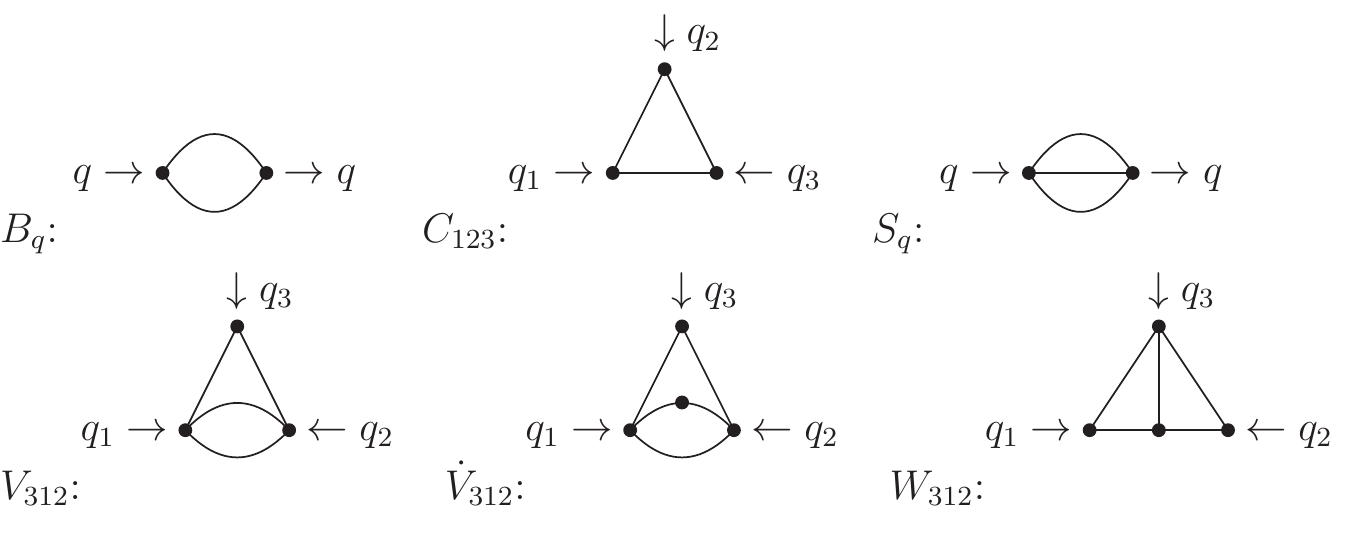}
\caption{\label{fig:masterint}Master integrals appearing in the two-loop calculation.  
The dot on the propagator in $\dot{V}_{312}$ corresponds to a doubling of that propagator.}
\end{figure}

The gluonic corrections to the quark loop are obtained by including two quark-gluon vertices from the Dyson series expansion in~(\ref{eq:hlbltensor}), i.e.
\begin{align}
\label{eq:backdysonqcd}
\Pi ^{\mu_{1} \mu_{2} \mu_{3} \mu_{4}}
&
=
-i\int\frac{d^4 q_{3}}{(2\pi)^4}  \left(\prod_{i=1}^{4}\int d^{4}x_{i}\, e^{-i q_{i} x_{i}}\right) 
\nonumber \\
&
\times 
\langle 0 | T\left(\prod_{j=1}^{4}J^{\mu_{j}}(x_{j}) \frac{1}{2}\int d^4 z_{1}\, d^4 z_{2} \,
i \mathcal{L}_{\textrm{int}}^{\textrm{qgq}}(z_{1}) \, i \mathcal{L}_{\textrm{int}}^{\textrm{qgq}}(z_{2})
\right) |0\rangle \, .
\end{align}
Denoting colour indices with bars, the interaction Lagrangians above are of the form
\begin{align}
\mathcal{L}_{\textrm{int}}^{\textrm{qgq}}(z_{i}) = g_{S}\frac{\lambda ^{a_{i}}_{\bar{\gamma }_{i}\bar{\delta}_{i}}}{2}
B^{a_{i}}_{\nu _{i}}(z_{i}) \, 
\bar{q}^{\bar{\gamma} _i}(z_i)\gamma ^{\nu _i}q^{\bar{\delta} _i}(z_i)  \, ,
\end{align}
where $B^{a_{i}}_{\nu _{i}}$ is the gluon field, $g_{S}$ is the strong coupling and $\lambda ^{a_{i}}$ is an $SU(3)_c$ Gell-Mann matrix. The only nonzero topology at this order is obtained by connecting all the quarks to the same line. Two examples of the diagrams in question are shown in Fig.~\ref{fig:2looptop}.
As a consequence of the topology, both the quark electric charge $e_{q}^4$ and the colour factor,
$\mathrm{Tr}(\lambda^{a}\lambda^{b})\delta_{ab}=2(N_{c}^2-1)$,
can be factored out, allowing to re-express the total contribution as a sum of all possible hexagons where two of the external lines are to be contracted to form the second loop, i.e.,
\begin{equation}
\Pi^{\mu_{1}\mu_{2}\mu_{3}\mu_{4}}=\frac{(N_{c}^2-1)g_{s}^2e_{q}^4}{4}\int \frac{d^{4}q_{5}}{(2\pi)^{4}}\frac{g_{\mu_{5}\mu_{6}}}{q_{5}^2}\lim_{ q_{6}\rightarrow -q_{5}}H^{\mu_{1}\mu_{2}\mu_{3}\mu_{4}\mu_{5}\mu_{6}} \, ,
\end{equation}
where
\begin{align}\nonumber
H^{\mu_{1}\mu_{2}\mu_{3}\mu_{4}\mu_{5}\mu_{6}}&\equiv\int \frac{d^4p}{(2\pi)^4} \sum_{\sigma(1,2,4,5,6)} \mathrm{Tr}\Bigg(\gamma^{\mu_{3}}S(p+q_1+q_2+q_4+q_5+q_6)\gamma^{\mu_1}S(p+q_2+q_4+q_5+q_6)\\&\times \gamma^{\mu_2}S(p+q_4+q_5+q_6)\gamma^{\mu_4}S(p+q_5+q_6)\gamma^{\mu_5}S(p+q_6)\gamma^{\mu_6}S(p)\Bigg)\, .
\end{align}
Here, $S(p)=\frac{\slashed{p}}{p^2}$ is the massless quark propagator and $\sigma (1,2,4,5,6)$ the set of pairwise permutations of $\mu_{i}$ and $q_{i}$ for $i=1,\, 2,\, 3, \, 5, \, 6$. 
The corresponding $\tilde{\Pi}_{i}$ for the two loops are
\begin{align}
\tilde{\Pi}_{i}&=P^{\tilde{\Pi}_{i}}_{\mu_1 \mu_2 \mu_3 \mu_4 \nu_4}\lim_{q_4\to 0} \frac{\partial \Pi^{\mu_{1} \mu_{2} \mu_{3} \nu_{4}}}{\partial q_{4}^{\mu_{4}}} \nonumber \\
&= -\frac{(N_{c}^2-1)g_{s}^2e_{q}^4}{4}\int \frac{d^{4}q_{5}}{(2\pi)^{4}}\frac{g_{\mu_{5}\mu_{6}}}{q_{5}^2}   \lim \limits_{\substack{
q_4 \to 0\\
q_6 \to -q_5}}P^{\tilde{\Pi}_{i}}_{\mu_1 \mu_2 \mu_3 \mu_4 \nu_4}\frac{\partial}{\partial q_{4}^{\nu_{4}}}H^{\mu_{1}\mu_{2}\mu_{3}\mu_{4}\mu_{5}\mu_{6}} \, .
\end{align}
After taking the derivative, using
\begin{equation}
\frac{\partial}{\partial q_{4}^{\nu_{4}}}S(p+q_{4})=-S(p+q_{4})\gamma_{\nu_{4}}S(p+q_{4}) \, ,
\end{equation}
the limit $q_4\to0$ and the projectors, we have for every $\tilde{\Pi}_{i}$ a large set of scalar two-loop integrals depending on two external momenta, $q_{1}$, and $q_{2}$, which can be parametrized as
\begin{align}
M(i_1,...,i_7) &= \frac{1}{i^2}\int \frac{d^d p_1}{(2\pi)^d}\int \frac{d^d p_2}{(2\pi)^d}\notag\\
&~~
\frac{1}{p_1^{2i_1}(p_1-q_1)^{2i_2}(p_1+q_2)^{2i_3}p_2^{2i_4}(p_2-q_1)^{2i_5}(p_2+q_2)^{2i_6}(p_1-p_2)^{2i_7}}\,.
\end{align}
Using {\sc KIRA} \cite{Maierhoefer:2017hyi} they can be reduced to the ones in Table \ref{tab:setint}, whose corresponding topologies are represented in Fig. \ref{fig:masterint}. This reduction is done in $d=4-2\epsilon\ne4$ dimensions.

\begin{table}[tb]
\begin{center}
\begin{tabular}{|cc|}
\hline
$i_1,...,i_7$ & $M(i_1,...,i_7)$ \\
\hline
$1,1,0,1,1,0,0$ & $B_1^2$\\
$1,0,1,1,0,1,0$ & $B_2^2$\\
$0,1,1,0,1,1,0$ & $B_3^2$\\
$1,0,1,1,1,0,0$ & $B_1 B_2$\\
$0,1,1,1,1,0,0$ & $B_1 B_3$\\
$0,1,1,1,0,1,0$ & $B_2 B_3$\\
$1,1,1,1,1,0,0$ & $B_1 C_{123}$\\
$1,1,1,1,0,1,0$ & $B_2 C_{123}$\\
$1,1,1,0,1,1,0$ & $B_3 C_{123}$\\
$0,1,0,1,0,0,1$ & $S_1$\\
$0,0,1,1,0,0,1$ & $S_2$\\
$0,0,1,0,1,0,1$ & $S_3$\\
$0,0,1,1,1,0,1$ & $V_{123}$\\
$0,0,1,1,1,0,2$ & $\dot V_{123}$\\
$1,0,1,0,1,0,1$ & $V_{213}$\\
$1,0,1,0,1,0,2$ & $\dot V_{213}$\\
$0,1,1,1,0,0,1$ & $V_{312}$\\
$0,1,1,1,0,0,2$ & $\dot V_{312}$\\
$0,1,1,1,0,1,1$ & $W_{123}$\\
$0,1,1,1,1,0,1$ & $W_{213}$\\
$1,0,1,1,1,0,1$ & $W_{312}$\\
$1,1,1,1,1,1,0$ & $C_{123}^2$\\
\hline  
\end{tabular}
\end{center}
\caption{List of master integrals $M(i_1,...,i_7)$ needed for the massless fully off-shell triangle at two-loop order. The last one is not needed at this
order.\label{tab:setint}}
\end{table}

This is a good point to discuss how we handle renomalization and regularization. Both ultraviolet and infrared divergences are regulated using dimensional regularization. We work to the lowest order in $\alpha$ and to first order in $\alpha_S$ in the massless quark limit. There are no
counterterms needed to this order and infrared divergences must vanish since the three photon 
amplitude vanishes because of charge-conjugation. However, individual diagrams and master integral
can be infrared and ultraviolet divergent. The quantities $\tilde\Pi_i$ are finite and the cancellation of all divergences, up to $1/\epsilon^3$ provides another good check on our calculations.

Strong efforts have been made to successfully obtain compact analytical expressions for all those two-loop integrals. All of the appearing master integrals can be found in terms of classical polylogarithms in Refs.~\cite{Birthwright:2004kk,Chavez:2012kn} up to the order that we need. They are collected together with their corresponding $\epsilon$ expansions in App.~\ref{app:mi}. Using these we find, as expected, that all the intermediate divergences exactly cancel, leading to a result of the following form
\begin{align}\nonumber
\tilde{\Pi}_{m}&=f_{m,ijk}^{pqr}F_{ijk}(2)Q_1^{2p}Q_2^{2q}Q_3^{2r}+w_{m,ijk}^{pqr}W_{ijk}(0)Q_1^{2p}Q_2^{2q}Q_3^{2r}+c_{m,ijk}^{pqr}C_{ijk}(0)Q_1^{2p}Q_2^{2q}Q_3^{2r}
\\
&+n_{m,1}^{pqr}Q_1^{2p}Q_2^{2q}Q_3^{2r}\log\frac{Q_{1}^2}{Q_{3}^2}+n_{m,2}^{pqr}Q_1^{2p}Q_2^{2q}Q_3^{2r}C_{ijk}(0)\log\frac{Q_{2}^2}{Q_{3}^2} \nonumber
\\
&+l_{m,ijk1}^{pqr}Q_1^{2p}Q_2^{2q}Q_3^{2r}C_{ijk}(0)\log\frac{Q_{1}^2}{Q_{3}^2}+l_{m,ijk2}Q_1^{2p}Q_2^{2q}Q_3^{2r}C_{ijk}(0)\log\frac{Q_{2}^2}{Q_{3}^2}\, . 
\end{align}
The remaining loop functions, $C_{ijk}(0),W_{ijk}(0),F_{ijk}(2)$ can also be found in App. \ref{app:mi}.
The explicit numerical coefficients can be found in the file {\tt pitildes.txt} of the supplementary material. In Table \ref{tab:pitildenums} we give numerical results for them in a benchmark point, $(Q_{1}^2,Q_{2}^2,Q_{3}^2)=(1,1.3,1.7)\,$ GeV$^2$, giving also the analogous quark loop ones for comparison. The loop corrections are found of the order of $\sim -\frac{\alpha_{s}}{\pi}$, i.e. they are found to be small as far as $\alpha_{s}$ is not large. The scale at which $\alpha_{s}$ should be set is similar to the scale $Q^{2}$ at which the $\tilde{\Pi}$ are evaluated. Otherwise, large logarithms $\ln\frac{\mu}{Q_{i}}$ appearing at higher orders would break the perturbative series. As a consequence, the series are found to be reliable as far as we do not go below $\sim 1 \,\mathrm{GeV}$.

Taking the linear combinations of the $\tilde{\Pi}_{i}$ which lead to the $\hat{\Pi}$, one finds analogous expressions for them, but with explicit negative powers of K\"all\'{e}n functions $\lambda= (Q_1^2+Q_2^2-Q_3^2)^2-4Q_1^2 Q_2^2$. They introduce singularities which are, however, spurious. When expanding around them, they cancel against the zeros of the polylogarithms, as explicitly checked in different kinematic limits. Details on these expansions can be found in App.~\ref{app:mi}. Numerical values for the $\hat{\Pi}_i$ in the same benchmark point and comparison with the corresponding quark loop are given in Table \ref{tab:pihatnums}. The analytical expressions are too long to be included here and are given in the file {\tt resultsgluon.txt} of the supplementary material. The equivalent results for the massless quark loop are in the file {\tt resultsquark.txt}. We have, however, included analytical expressions for both the quark loop and gluonic correction at the symmetric point $Q_{1}=Q_{2}=Q_{3}$ in App.~\ref{app:analytic}.

\begin{table}[t!]\centering
\begin{tabular}{|c|r|r|r|r|r|}
\hline
\rule{0cm}{14pt}& $\tilde{\Pi}_{1}$ & $\tilde{\Pi}_{7}$ & $\tilde{\Pi}_{10}$ & $\tilde{\Pi}_{13}$ & $\tilde{\Pi}_{19}$ \\ \hline
Quark loop                                            & $-0.0816$        & $0.123$          & $0.0363$           & $0.0274$           & $0.0263$           \\ \hline
Gluon corrections ($\times{\pi}/{\alpha_{s}}$) & $0.0781$         & $-0.136$         & $-0.0376$          & $-0.0398$          & $-0.0411$           \\ \hline
\end{tabular}
\caption{\label{tab:pitildenums}Values for the quark loop and gluonic correction contributions to the $\tilde{\Pi}$ in GeV units for a benchmark tuple $(Q_{1}^2,Q_{2}^2,Q_{3}^2)=(1,1.3,1.7)\,$ GeV$^2$. Sum over the three flavours has been made. The last line is in units of ${\alpha_{s}}/{\pi}$.}
\end{table}

\begin{table}[t!]\centering
\begin{tabular}{|c|r|r|r|r|r|r|}
\hline
\rule{0cm}{14pt}& $\hat{\Pi}_{1}$ & $\hat{\Pi}_{4}$ & $\hat{\Pi}_{7}$ & $\hat{\Pi}_{17}$ & $\hat{\Pi}_{39}$ &$\hat{\Pi}_{54}$ \\ \hline
Quark loop                                            & $-0.0210$        & $-0.0119$          & $-0.00384$           & $0.00386$           & $0.0119$    & $0.000422$        \\ \hline
Gluon corrections ($\times\pi/{\alpha_{s}}$) & $0.0178$        & $0.00560$          & $0.00302$           & $-0.00750$           & $-0.0103$    & $-0.000427$       \\ \hline
\end{tabular}
\caption{\label{tab:pihatnums}Values for the quark loop and gluonic correction contributions to the $\hat{\Pi}$ in GeV units for a benchmark tuple $(Q_{1}^2,Q_{2}^2,Q_{3}^2)=(1,1.3,1.7)\,$ GeV$^2$. Sum over the three flavours has been made. The last line is in units of ${\alpha_{s}}/{\pi}$.}
\end{table}

A possible check on our result is taking the limit where one of the virtualities is much smaller than the other two, i.e. the limit of Ref.~\cite{Melnikov:2003xd}, where it is argued that the leading term should have no corrections. The consequences of this limit for the $\hat \Pi_i$ has been analyzed in Ref.~\cite{Colangelo:2019uex}, where it is shown that only for $\hat\Pi_1$ there is an unambiguous prediction. Taking into account the corrections to the OPE of two-photon currents to the axial current, i.e. the axial current gets an extra factor of $1-\alpha_S/\pi$~\cite{Ludtke:2020moa,Kodaira:1978sh,Kodaira:1979ib,Kodaira:1979pa}, we see that our result indeed satisfies the arguments of  Ref.~\cite{Melnikov:2003xd}.

\section{Results for the $(g-2)_{\mu}$ and phenomenological implications}
\label{sec:numerics}
Now that we have the needed gluonic corrections to the $\hat{\Pi}_i$, we can introduce them into~(\ref{eq:amuint}) to calculate their corresponding contributions to $a_{\mu}^{\mathrm{HLbL}}$. Obviously, the identification of the $\hat{\Pi}_i$ with the ones obtained using the OPE only makes sense when such an expansion is valid, i.e., above some cut for the Euclidean momenta, $Q_{1,2,3}>Q_{\mathrm{min}}$. We restrict ourselves to those integration regions, keeping in mind that the (dominant) contributions from the remaining regions, necessarily computed with non-perturbative methods, must be added to the ones computed here.

The numerical integration has been done with the {\sc VEGAS} implementation in
the {\sc CUBA} library, as well as our own implementation of two deterministic algorithms. We have checked that the results agree. The general expressions for the quark loop and the gluonic corrections have large negative powers of $\lambda$ and become numerically unstable whenever $\lambda$ is small. We therefore use, as in our previous work for the quark loop~\cite{Bijnens:2020xnl}, expansions whenever that happens. There are six different expansions that need to be done. This is explained in more detail in App.~\ref{app:analytic}. We have checked that the numerical results are not sensitive to changing the boundaries where we use the different expansions.

We perform the integrals of the $12\, \bar{\Pi}_i$ contributions at different $Q_{\mathrm{min}}$, both for the leading OPE contribution, the quark loop, and the gluonic corrections. They are displayed for $Q_{\mathrm{min}}=1$~GeV in Table \ref{tab:intQ1}.

\begin{table}[t!]\centering
\begin{tabular}{|l|r|r|}
\hline        & Quark loop & Gluon corrections ($\frac{\alpha_{s}}{\pi}$ units) \\ \hline
$\bar{\Pi}_{1}$  & $0.0084$   & $-0.0077$                                                                          \\ \hline
$\bar{\Pi}_{2}$  & $13.28$  & $-12.30$                                                                       \\ \hline
$\bar{\Pi}_{3}$  & $0.78$     & $-0.87$                                                                          \\ \hline
$\bar{\Pi}_{4}$  & $-2.25$    & $0.62$                                                                 \\ \hline
$\bar{\Pi}_{5}$  & $0.00$     & $0.20$                                                     \\ \hline
$\bar{\Pi}_{6}$  & $2.34$     & $-1.43$                                                                           \\ \hline
$\bar{\Pi}_{7}$  & $-0.097$   & $0.056$                                                                            \\ \hline
$\bar{\Pi}_{8}$  & $0.035$    & $0.41$                                                                         \\ \hline
$\bar{\Pi}_{9}$  & $0.623$    & $-0.87$                                                                           \\ \hline
$\bar{\Pi}_{10}$ & $1.72$     & $-1.61$                                                                             \\ \hline
$\bar{\Pi}_{11}$ & $0.696$   & $-1.04$                                                                            \\ \hline
$\bar{\Pi}_{12}$ & $0.165$    & $-0.16$                                                                            \\ \hline
Total            & $17.3$     & $-17.0$                                                                      \\ \hline
\end{tabular}
\caption{Leading contributions to the $(g-2)_{\mu}$ integration from $Q_{\mathrm{min}}=1\, \mathrm{GeV}$ in $10^{-11}$ units.}
\label{tab:intQ1}
\end{table}

Consistently with the size of the gluonic corrections found for the $\tilde{\Pi}_i$ in the previous section, we find that they are negative and of order $\frac{\alpha_{s}}{\pi}$.
Given the power fall-off of the contributions of the $\hat{\Pi}$ with respect to the studied energies, the quantitative contribution above some energy cut $Q_{\mathrm{min}}$ is saturated by the regions nearby such a cut. As a consequence, a natural scale to effectively avoid large logarithms in the corresponding perturbative series is $\mu\sim 
Q_{\mathrm{min}}$, however the exact choice of it is ambiguous. In order to estimate perturbative uncertainties we will vary the scale dependence, a consequence of cutting the series at two-loops or at the first $\alpha_S$ correction, in the interval $\mu^2\in(\frac{1}{2},2)Q^2_{\mathrm{min}}$. At the studied order, the whole scale dependence comes from $\alpha_{s}(\mu)$. Taking $\alpha_{s}^{N_{f}=5}(M_{Z})$ from Ref. \cite{Aoki:2019cca}, we run it at five loops to $\alpha_{s}^{N_{f}=3}(m_{\tau})$.\footnote{We implement the running, in the conventional $\overline{\mathrm{MS}}$ scheme, using {\sc Rundec} \cite{Herren:2017osy}.} As a further conservative estimates of perturbative uncertainty, we add quadratically the difference obtained by taking the one obtained running from $\alpha_{s}^{N_{f}=3}(m_{\tau})$ to $\alpha_{s}^{N_{f}=3}(Q_{\mathrm{min}})$ with the five loop running (which we take as our central result) with the one obtained keeping $\alpha_{s}^{N_{f}=3}$ with a fixed scale, $\mu=m_{\tau}$, which at the order we are working with is also a legitimate choice. Finally we also add quadratically the subleading uncertainty coming from $\alpha_{s}^{N_{f}=5}(M_{Z})$.

\begin{figure}[tb]\centering
\includegraphics[width=0.85\textwidth]{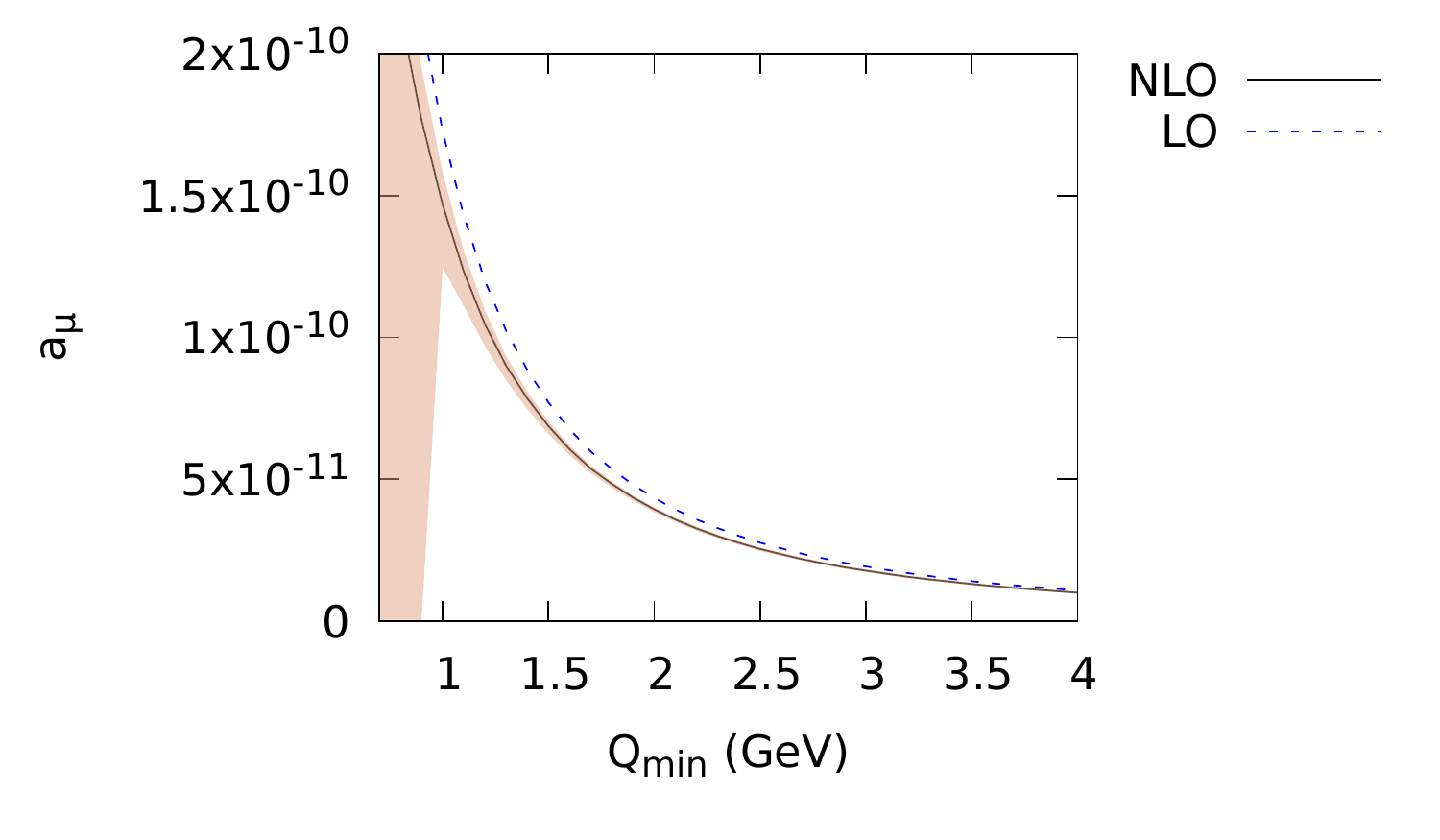}
\caption{\label{fig:plotnlo}Numerical results for 
the hadronic HLbL $(g-2)_{\mu}$ in the $Q_{i}>Q_{\mathrm{min}}$ region, using the LO (massless quark loop) and NLO (gluonic corrections) contributions of its corresponding OPE. Uncertainties, apart from the one coming from the $\alpha_{s}(M_{W})$ input, represented by shaded areas have being estimated attending to ambiguities when setting the $\alpha_{s}$ input (exact choice of scale and order of running for the $\beta$ function) as a consequence of not including higher-orders.}
\end{figure}

The result, where we show the quark loop, the gluonic corrections and the obtained uncertainties is shown in Figure \ref{fig:plotnlo}. While in general we consider our uncertainty estimates reliable, we notice that they may be slightly over-conservative in the region just below $1\, \mathrm{GeV}$ and over-optimistic just above it. This is a consequence of the sharp break down of the $\alpha_{s}$ running at $\mu\sim 0.7 \, \mathrm{GeV}$ which makes our uncertainty strongly dependent on the exact scale interval chosen to estimate them. In essence, we find that the correction is small and negative and that the series are well-behaved, having a gluonic correction of around $-10 \%$ above the perturbative breakdown.

\section{Conclusions}\label{sec:concl}
One of the main sources of uncertainties entering in $(g-2)_{\mu}$ comes from the contributions of the short-distance regions of the HLbL tensor contributions. In this work, which can be regarded as a continuation of Refs. \cite{Bijnens:2019ghy} and \cite{Bijnens:2020xnl}, we have culminated our task of giving a precise and systematic description of the contributions for three large loop momenta. 

For years, it was assumed that some form of the quark loop, maybe with constituent quark masses, should be the leading order of some systematic expansion of the HLbL contribution tensor to the $(g-2)_{\mu}$ for large loop momenta. However, it was shown in Ref. \cite{Bijnens:2019ghy} how applying an OPE directly to the HLbL tensor, where the massless quark loop is indeed the leading order, does not make sense for the $(g-2)_{\mu}$ kinematics. The correct expansion in this kinematic region was presented in that reference, where the massless quark loop was shown to be the leading order and the leading non-perturbative quark mass-suppressed correction was computed. 

A very comprehensive analysis to study the role of both the quark mass-suppressed and not suppressed non-perturbative corrections to the expansion was made in Ref.~\cite{Bijnens:2020xnl}, where many formal aspects and subtleties of the expansion were developed and presented in full detail, showing that it is well founded. The obtained results showed how above $1 \, \mathrm{GeV}$ the non-perturbative corrections, even when functionally more important than in other expansions, are still typically below $1\%$.

In view of that, the most important corrections to the leading massless quark loop, and the one that ultimately allows to understand from where the expansion is valid, is the pure gluonic correction, which has been the subject of this work. 

While in principle a multi-scale four-loop integral could be regarded as a formidable task, it has become feasible through combining existing tools developed for generic contributions of the HLbL tensor to the $(g-2)_{\mu}$, methods on finding compact expressions developed in Ref.~\cite{Bijnens:2020xnl}, optimized software on reduction to master integrals, analytic reduction of those remaining master integrals and numerical integration routines.

Our final result brings good news. The size of the gluonic corrections are found small, typically of size $-10 \%$ above the perturbative breakdown scale, and, as a consequence, the expansion is able to give a precise description of the $(g-2)_{\mu}$ contributions above it. 

Taking all of this into account, we suggest as a legitimate method to compute the HLbL contribution to $(g-2)_{\mu}$ to use the results of this expansion from some point between $Q_{\mathrm{min}}=1$ GeV and $Q_{\mathrm{min}}=2$ GeV, which should give a more precise prediction than resonance models, and possible discontinuities in the matching should be incorporated as systematic model uncertainty.


\section*{Acknowledgments}
We thank Martin Hoferichter for discussions.
N.~H.--T. and L.L.~are funded by the Albert Einstein Center for Fundamental Physics at Universit\"{a}t Bern and the Swiss National Science Foundation, respectively. J.~B. is supported in part by the Swedish Research Council grants contract numbers 2016-05996 and 2019-03779. A. R-.S- is partially supported by the Agence Nationale de la Recherche (ANR) under grant ANR-19-
CE31-0012 (project MORA).



\appendix
\renewcommand{\theequation}{\thesection.\arabic{equation}}

\section{Master integrals}\label{app:mi}

The aim of this appendix is to list the expressions of the master integrals needed in Sec.~\ref{sec:2loop} (see also Fig.~\ref{fig:masterint}). They can be found in Refs.~\cite{Usyukina:1994iw,Birthwright:2004kk,Chavez:2012kn} \footnote{The formulas given in this appendix and those in Ref.~\cite{Chavez:2012kn} differ in a sign: an overall minus sign has been missed in (4.18) of Ref.~\cite{Chavez:2012kn}. This in turns leads to a minus instead of the plus sign in the second line of (4.24), which corresponds to our (\ref{eq:Vdotepsexp}). We checked that our sign agrees with the corresponding formulas in Ref.~\cite{Usyukina:1994iw}, which is also cited in Ref.~\cite{Chavez:2012kn}.}. All n-loop master integrals contain the overall factor $S_D^n$, where
\begin{equation}
    S_D=S_D(\eps)=\frac{(4\pi)^\eps}{16 \pi ^2}\frac{\Gamma (1+\eps)\Gamma ^2(1-\eps)}{\Gamma (1-2\eps)}\,.
\end{equation}
The functions $C_{123}$ and $W_{312}$ are finite in the limit $d\to4$. Their $\eps$-expansions can be written as follows
\begin{align}\label{eq:masterintegralC}
   C_{123}\equiv & \int dp\frac{1}{p^{2\gamma}(p-q_1)^{2}(p+q_2)^{2}} \nonumber \\
   =&\, S_D C_{123}(0)+S_D C_{123}(1)\eps +C_{123}(2)\eps ^2+\mathcal{O} (\eps^3)\,,
\end{align}
\begin{align}\label{eq:masterintegralW}
   W_{312} \equiv & \int dp_1 dp_2 \frac{1}{p_1^2(p_1+q_2)^2 p_2^2 (p_2-q_1)^2 (p_1-p_2)^2} \nonumber \\
   =& \, S_D^2 W_{312}(0)+S_D^2 W_{312}(1)\eps + W_{312}(2)\eps^2+\mathcal{O} (\eps^2)\,,
\end{align}
where $q_1+q_2+q_3=0$ and $d=4-2\eps$. The integrations are defined as
\begin{equation}
    \int dp = \frac{1}{i}\int \frac{d^dp}{(2\pi)^d}
    \, .
\end{equation}
The remaining integrals appearing in the calculation contain some singularities which cancel out in the final amplitude. The $\eps$-expansions of the integral $B_q$ and $S_q$ read
\begin{align}
    B_q&\equiv \int dp \frac{1}{p^{2}(p-q)^{2}} \nonumber \\
    &=\frac{S_D}{\epsilon}+S_D\left[2-\log (-q^2)\right]+S_D\left[ 4-2\log(-q^2)+\frac{1}{2}\log^2(-q^2)\right]\epsilon+\mathcal{O}(\eps ^2)\,,
\end{align}
and
\begin{align}
    S_q\equiv &\int dp_1 dp_2 \frac{1}{(p_1-q)^2 p_2^2 (p_1-p_2)^2} \nonumber \\
        =&-S_D^2 \frac{q^2}{4\eps}
       + S_D^2 \left[  - \frac{13}{8}q^2 + \frac{1}{2}\log( - q^2) q^2 \right] \nonumber \\
       &+S_D^2 \left[  - \frac{115}{16}q^2 + \frac{13}{4}\log( - q^2) q^2 - \frac{1}{2}\log^2( -q^2)q^2 \right]\eps+\mathcal{O}(\eps^2)\,. 
\end{align}
The integrals $V_{123}$ and $\dot{V}_{123}$ can be expressed as functions of the finite integrals $C_{123}$ and $W_{312}$. Their $\eps$-expansions can be written as
\begin{align}\label{eq:Vdotepsexp}
    &\dot{V}_{312}\equiv \int dp_1 dp_2 \frac{1}{p_1^2(p_1-q_1)^2 (p_1+q_2)^2 p_2^2 (p_1-p_2)^2} \nonumber \\
    &=-S_D^2\frac{ C_{123}(0)}{\eps}+S_D^2\left[\frac{1}{2}C_{123}(0)\left( \log (-q_1^2)+\log(-q_2^2)\right)-C_{123}(1)\right] \nonumber \\
    &- \frac{S_D^2}{4} \left[C_{123}(0) \left(\log ^2(-q_1^2)+\log ^2(-q_2^2)\right)-2 C_{123}(1) \left( \log (-q_1^2)+\log(-q_2^2)\right)\right. \nonumber \\
    &\qquad \, \, \, \, \left. +4 (C_{123}(2)+W_{312}(0))\right]\eps+\mathcal{O}\left( \eps ^2\right)\,,
\end{align}
    and 
\begin{align}
    &V_{312}\equiv \int dp_1 dp_2 \frac{1}{(p_1-q_1)^2 (p_1+q_2)^2 p_2^2 (p_1-p_2)^2} \nonumber \\
    &=\frac{S_D^2}{2\, \eps ^2}+S_D^2 \left[\frac{5}{2}-\log(-q_3^2)\right] \frac{1}{\eps} +\frac{S_D^2}{2}\left[ C_{123}(0) (-q_1^2-q_2^2+q_3^2)\right. \nonumber\\
    &\qquad \left. +\log (-q_1^2) \left(\log (-q_3^2)-\log (-q_2^2)\right) + \log (-q_2^2) \log (-q_3^2)+\log ^2(-q_3^2)-10 \log (-q_3^2)+19 \right]\nonumber \\
    &+S_D^2 \left[ \frac{1}{2} \left(F_{312}(2)+65\right)+\frac{1}{2} C_{123}(1) (-q_1^2-q_2^2+q_3^2)\right. \nonumber \\
    &\qquad +C_{123}(0) \left(-\frac{5}{2} (q_1^2+q_2^2-q_3^2)+\frac{1}{4} \log(-q_1^2) (q_1^2+q_2^2-q_3^2)+\frac{1}{4} \log (-q_2^2) (q_1^2+q_2^2-q_3^2)\right) \nonumber \\
   &\qquad +\log (-q_1^2) \left(\log (-q_2^2) \left(\log
   (-q_3^2)-\frac{5}{2}\right)-\log ^2(-q_3^2)+\frac{5 \log (-q_3^2)}{2}\right)\nonumber \\
   &\qquad \left.+\log (-q_2^2) \left(\frac{5 \log (-q_3^2)}{2}-\log ^2(-q_3^2)\right)+\frac{1}{3} \log
   ^3(-q_3^2)+\frac{5}{2} \log ^2(-q_3^2)-19 \log (-q_3^2)\right]\eps \nonumber\\
   &+\mathcal{O}(\eps ^2)\,,
   %
\end{align}
where $C_{123}(i)$ and $W_{312}(i)$ are the coefficients of the $\eps$-expansion of their corresponding Master integral (c.f.  (\ref{eq:masterintegralC}) and (\ref{eq:masterintegralW})). Not all these coefficients survive in the gluonic corrections we are computing in section \ref{sec:2loop}: $C_{123}(1)$, $C_{123}(2)$, $W_{312}(1)$ and $W_{312}(2)$ cancel in the final expression. The coefficients $C_{123}(0)$ and $W_{312}(0)$ that contribute, as well as the function $F_{312}(2)$ which appears in the expansion of $V_{312}$, are given below

\begin{equation}
    C_{123}(0)=2q_3^{-2}\frac{\mathcal{P}_2(z)}{z-\overline{z}}\,,
\end{equation}
\begin{equation}
    W_{312}(0)=6q_3^{-2}\frac{\mathcal{P}_4(1-z^{-1})}{z-\overline{z}}\,,
\end{equation}
and finally
\begin{equation}
    F_{312}(2)= -6\mathcal{P}_3(z)-6\mathcal{P}_3(1-z)+\frac{1}{2}\log(u)\log ^2(v)+\frac{1}{2}\log ^2(u)\log (v)+6 \zeta _3\,,
\end{equation}
where $u=\frac{q_1^2}{q_3^2}$, $v=\frac{q_2^2}{q_3^2}$ and $z$ is given by
\begin{align}
z &=\frac{1}{2}\left(1+u-v+i\sqrt{-\bar\lambda}\right)\,,
\notag\\
\bar\lambda&= (1+u-v)^2-4u\,.
\end{align}
The $\mathcal{P}_i(z)$ are real (purely imaginary) functions over the complex plane when i is odd (real). They can be expressed using polylogarithms:
\begin{align}\label{eq:Pidefs}
\mathcal{P}_2(z) &= \Li_2(z)-\Li_2(\bar z)+\log|z|(\log(1-z)-\log(1-\bar z))
\nonumber \\
\mathcal{P}_3(z) &= \Li_3(z)+\Li_3(\bar z)-\log|z|(\Li_2(z)+\Li_2(\bar z))-\frac{1}{3}\log^2|z|(\log(1-z)+\log(1-\bar z))
\nonumber \\
\mathcal{P}_4(z) &= \Li_4(z)-\Li_4(\bar z)-\log|z|(\Li_3(z)-\Li_3(\bar z))+\frac{1}{3}\log^2|z|(\Li_2(z)-\Li_2(\bar z))
\end{align}
The polylogarithms can be defined recursively
\begin{equation}
    \Li_n(z)=\int_0^z \frac{dt}{t}\Li _{n-1}(t) \text{ , and }\Li_1(z)=-\log(1-z)\,.
\end{equation}
The $\mathcal{P}_i$ satisfy a number of relations
\begin{align}\label{eq:Pirels}
    \mathcal{P}_2(z)&=-\mathcal{P}_2(1/z)\,,
    \nonumber \\
    \mathcal{P}_3(z)&=\mathcal{P}_3(1/z)\,,
    \nonumber \\
    \mathcal{P}_4(z)&=-\mathcal{P}_4(1/z)\,,
    \nonumber \\
    \mathcal{P}_2(z)&=\mathcal{P}_2(1-1/z)=-\mathcal{P}_2(1-z)=\mathcal{P}_2(1/(1-z))=-\mathcal{P}_2(z/(z-1))\,,
    \nonumber \\
    \mathcal{P}_3(z)&+\mathcal{P}_3(1-z)+
    \mathcal{P}_3(1-1/z)=\mathcal{P}_3(1)=2\zeta(3)\,
\end{align}
which can be used to show that the master integrals have the required symmetries under interchange of momenta.

\section{Analytical formulae}\label{app:analytic}

In this section we present analytical formulae for the scalar functions entering into the calculation of $a_{\mu}^{\mathrm{HLbL}}$. We in particular discuss the momentum expansions of the master integrals needed to make spurious singularities cancel numerically. As an explicit example, we also give the expressions for the $\hat{\Pi}$ at the symmetric point $Q_1=Q_2=Q_3$ for the quark loop and gluonic correction in App.~\ref{app:symmetricpoint}.

\subsection{Expansions}
\begin{figure}[tb]\centering
\includegraphics[width=0.6\textwidth]{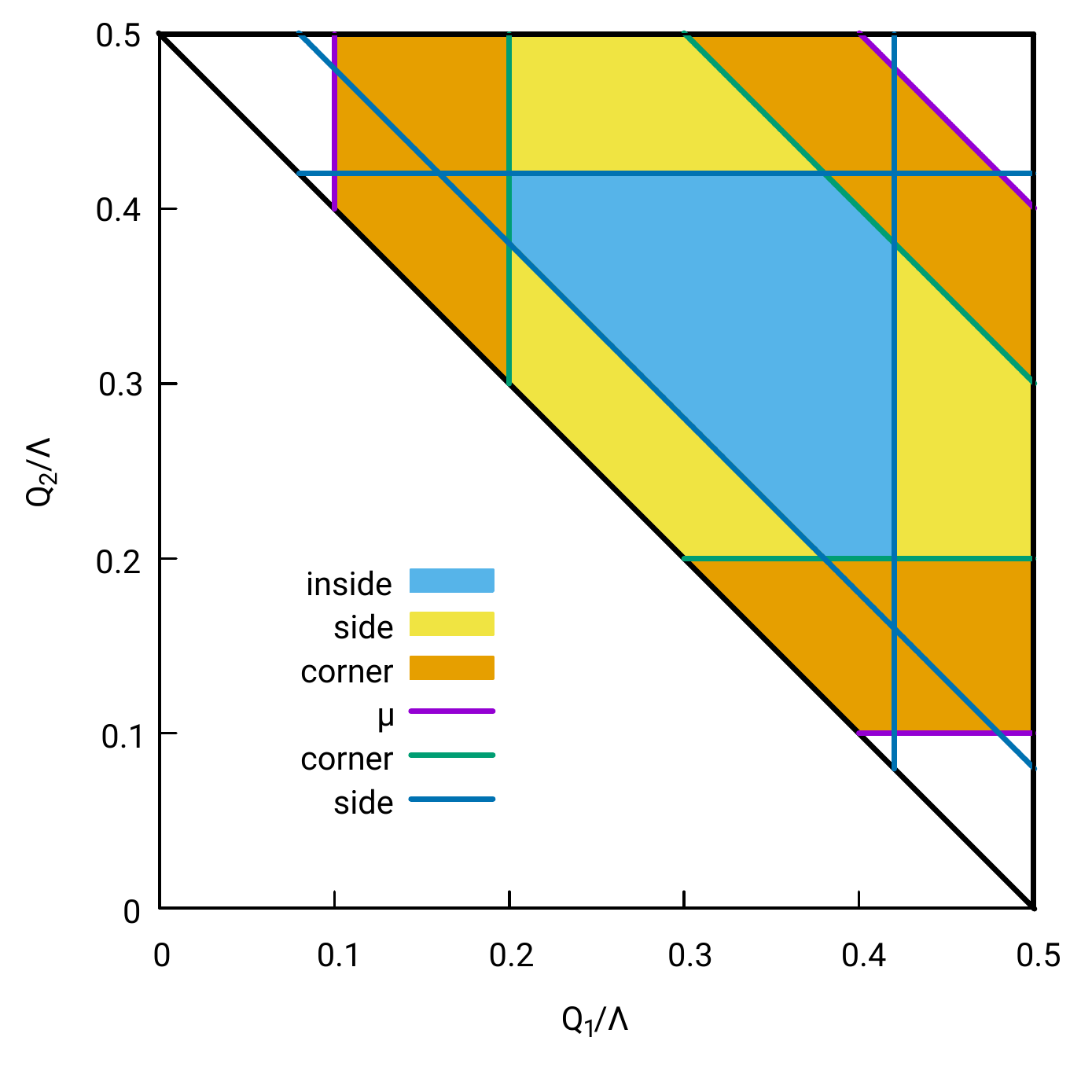}
\caption{\label{fig:triangleKallen}Different regions to consider in order to deal with the singularities of the $\hat{\Pi}$ when $\lambda \to 0$. The regions are shown for $Q_1+Q_2+Q_3=\Lambda$.}
\end{figure}

In the numerical evaluation of $a_{\mu}^{\mathrm{HLbL}}$ there are certain limits of the kinematics requiring particular care. The integration domain can be divided into several regions as in Fig.~\ref{fig:triangleKallen}. We there see the so-called side, corner and inside regions together with their boundaries. Also the cut-off $\mu$ has been indicated. Unless the side and corner regions are properly taken care of, the numerical integration will diverge as one obtains zeros in denominators that numerically do not cancel the zeros in numerators. Below we discuss the two types of problematic regions.

The precise definition of the regions is: $Q_i\ge \mu=Q_\textrm{min}$.
The corners are defined by $Q_i/(Q_1+Q_2+Q_3)\le \epsilon_1$ for $i=1,2,3$. The sides are the part of the remaining region that satisfy   $(2Q_i/(Q_1+Q_2+Q_3)-1\le \epsilon_2$ for $i=1,2,3$. The inside is the remaining allowed region.

\subsubsection{Side regions}

The side regions are defined as the kinematical limit where one $Q_{i}$ is close to $Q_{j}+Q_{k}$, or, in other words when
\begin{align}\label{eq:numside}
\textrm{Side Region }S_{i}: \; \; \; Q_{i}^2 = \left( Q_{j}+ Q_{k} \right) ^2 -\delta \equiv \overline{Q}_{i}^2-\delta  \, ,
\end{align}
where $\delta$ is a small parameter. The inverse powers of the K\"all\'en function in the $\hat{\Pi}_{i}$ diverge in the side regions. These apparent singularities do, however, cancel when all the kinematical factors, the master integrals $C_{123}(0)$, $W_{312}(0)$, $W_{213}(0)$, $W_{123}(0)$, $F_{312}(2)$, $F_{213}(2)$ and $F_{123}(2)$ as well as the K\"all\'en function itself are expanded in $\delta$. For a finite result we have to expand the master integrals up to order $\delta^9$. The analytical forms of these expansions are very long and we here therefore only give the first two orders for one case, $S_{3}$. In the supplementary file {\tt sideexpansions.txt}, however, we provide the full expansions needed for all $S_{ i}$. In region $S_{+3}$ we have

\begin{align}
C_{123}(0) =&\frac{1}{Q_2 \overline{Q}_3} \log \left( \frac{Q_1^2}{\overline{Q}_3^2}\right) + \frac{1}{Q_1 \overline{Q}_3} \log \left( \frac{Q_2^2}{\overline{Q}_3^2}\right) \nonumber\\
&+  \delta \Bigg[ \frac{Q_1}{6Q_2^2 \overline{Q}_3^3} \log \left( \frac{Q_1^2}{\overline{Q}_3^2}\right) 
+ \frac{Q_2}{6Q_1^2 \overline{Q}_3^3} \log \left( \frac{Q_2^2}{\overline{Q}_3^2}\right) 
+ \frac{1}{3Q_2 \overline{Q}_3^3} 
+ \frac{1}{2Q_2 \overline{Q}_3^3} \log \left( \frac{Q_1^2}{\overline{Q}_3^2}\right) \nonumber \\
& + \frac{1}{3Q_1 \overline{Q}_3^3}
+ \frac{1}{2Q_1 \overline{Q}_3^3} \log \left( \frac{Q_2^2}{\overline{Q}_3^2}\right) \Bigg] + \mathcal{O}(\delta ^2)\, , 
\end{align}
\begin{align}
F_{312}(2) =&- 6\zeta_3 + \frac{1}{2} \log \left( \frac{Q_1^2}{\overline{Q}_3^2}\right) \log ^2 \left( \frac{Q_2^2}{\overline{Q}_3^2}\right) + \frac{1}{2} \log ^2 \left( \frac{Q_1^2}{\overline{Q}_3^2}\right) \log \left( \frac{Q_2^2}{\overline{Q}_3^2}\right) + 6 \mathcal{P}_3\left( - \frac{Q_2}{Q_1}\right) \nonumber \\
&+  \delta \Bigg[  - \frac{Q_2}{Q_1 \overline{Q}_3^2} \log ^2 \left( \frac{Q_2^2}{\overline{Q}_3^2}\right) 
- \frac{Q_2}{Q_1 \overline{Q}_3^2} \log \left( \frac{Q_1^2}{\overline{Q}_3^2}\right) \log \left( \frac{Q_2^2}{\overline{Q}_3^2}\right)
- \frac{Q_2}{Q_1 \overline{Q}_3^2} \log ^2 \left( \frac{Q_1^2}{\overline{Q}_3^2}\right) \nonumber\\
&- \frac{3}{Q_1 \overline{Q}_3} \log \left( \frac{Q_2^2}{\overline{Q}_3^2}\right) 
+ \frac{1}{Q_1 \overline{Q}_3} \log ^2 \left( \frac{Q_2^2}{\overline{Q}_3^2}\right)
+  \frac{3}{Q_1 \overline{Q}_3} \log \left( \frac{Q_1^2}{\overline{Q}_3^2}\right)\nonumber \\
&+ \frac{1}{Q_1 \overline{Q}_3} \log \left( \frac{Q_1^2}{\overline{Q}_3^2}\right) \log \left( \frac{Q_2^2}{\overline{Q}_3^2}\right) + \frac{1}{Q_1 \overline{Q}_3} \log ^2 \left( \frac{Q_1^2}{\overline{Q}_3^2}\right) 
- \frac{3}{Q_1 Q_2} \log \left( \frac{Q_1^2}{\overline{Q}_3^2}\right) \Bigg] \nonumber \\
&+ \mathcal{O}(\delta ^2) \, ,
\end{align}
\begin{align}
W_{312}(0) =& \frac{3}{Q_1 Q_2} \mathcal{P}_3\left( - \frac{Q_2}{Q_1}\right)\nonumber \\
&+  \delta \Bigg[  
- \frac{1}{12 Q_1^2 \overline{Q}_3^2} \log ^2 \left( \frac{Q_2^2}{\overline{Q}_3^2}\right) 
+ \frac{1}{6Q_1^2 \overline{Q}_3^2} \log \left( \frac{Q_1^2}{\overline{Q}_3^2}\right) \log \left( \frac{Q_2^2}{\overline{Q}_3^2}\right) - \frac{1}{12Q_1^2 \overline{Q}_3^2} \log ^2 \left( \frac{Q_1^2}{\overline{Q}_3^2}\right) \nonumber \\
&- \frac{1}{2Q_1^2 Q_2 \overline{Q}_3} \log \left( \frac{Q_2^2}{\overline{Q}_3^2}\right) 
+ \frac{1}{12 Q_1^2 Q_2 \overline{Q}_3} \log ^2 \left( \frac{Q_2^2}{\overline{Q}_3^2}\right)
+ \frac{1}{2Q_1^2 Q_2 \overline{Q}_3} \log \left( \frac{Q_1^2}{\overline{Q}_3^2}\right) \nonumber \\
&- \frac{1}{6Q_1^2 Q_2 \overline{Q}_3} \log \left( \frac{Q_1^2}{\overline{Q}_3^2}\right) \log \left( \frac{Q_2^2}{\overline{Q}_3^2}\right)
+ \frac{1}{12Q_1^2 Q_2 \overline{Q}_3} \log ^2 \left( \frac{Q_1^2}{\overline{Q}_3^2}\right) 
- \frac{1}{2 Q_1^2 Q_2^2} \log \left( \frac{Q_1^2}{\overline{Q}_3^2}\right) \nonumber \\
&+ \frac{1}{2Q_1^2 Q_2^2} \mathcal{P}_3\left( - \frac{Q_2}{Q_1}\right) \Bigg] + \mathcal{O}(\delta ^2)\, , 
\end{align}
\begin{align}
F_{213}(2) =&- 6 \zeta_3 - \log ^3 \left( \frac{Q_2^2}{\overline{Q}_3^2}\right) + \frac{3}{2} \log \left( \frac{Q_1^2}{\overline{Q}_3^2}\right) \log ^2 \left( \frac{Q_2^2}{\overline{Q}_3^2}\right) - \frac{1}{2} \log ^2 \left( \frac{Q_1^2}{\overline{Q}_3^2}\right) \log \left( \frac{Q_2^2}{\overline{Q}_3^2}\right)\nonumber \\
&+ 6 \mathcal{P}_3\left(1 + \frac{Q_2}{Q_1}\right) 
+ \delta \Bigg[ 3 \frac{Q_2}{Q_1 \overline{Q}_3^2} \log \left( \frac{Q_2^2}{\overline{Q}_3^2}\right) 
+ \frac{3Q_2}{2Q_1 \overline{Q}_3^2} \log ^2 \left( \frac{Q_2^2}{\overline{Q}_3^2}\right) \nonumber \\
&- \frac{3 Q_2}{Q_1 \overline{Q}_3^2} \log \left( \frac{Q_1^2}{\overline{Q}_3^2}\right) 
-\frac{5Q_2}{2Q_1 \overline{Q}_3^2} \log \left( \frac{Q_1^2}{\overline{Q}_3^2}\right) \log \left( \frac{Q_2^2}{\overline{Q}_3^2}\right) + \frac{Q_2}{Q_1 \overline{Q}_3^2} \log ^2 \left( \frac{Q_1^2}{\overline{Q}_3^2}\right) \nonumber\\
&- \frac{3}{2Q_1 \overline{Q}_3} \log ^2 \left( \frac{Q_2^2}{\overline{Q}_3^2}\right) 
+  \frac{3}{Q_1 \overline{Q}_3} \log \left( \frac{Q_1^2}{\overline{Q}_3^2}\right)
+ \frac{5}{2Q_1 \overline{Q}_3} \log \left( \frac{Q_1^2}{\overline{Q}_3^2}\right) \log \left( \frac{Q_2^2}{\overline{Q}_3^2}\right) \nonumber \\
&- \frac{1}{Q_1 \overline{Q}_3} \log ^2 \left( \frac{Q_1^2}{\overline{Q}_3^2}\right) \Bigg] + \mathcal{O}(\delta ^2)\, , 
\end{align}
\begin{align}
W_{213}(0) =&- \frac{3}{Q_1 \overline{Q}_3} \mathcal{P}_3\left(1 + \frac{Q_2}{Q_1}\right) \nonumber \\
&+  \delta \Bigg[  
-  \frac{Q_2}{2Q_1^2 \overline{Q}_3^3} \log \left( \frac{Q_2^2}{\overline{Q}_3^2}\right) 
+ \frac{Q_2}{2Q_1^2 \overline{Q}_3^3} \log \left( \frac{Q_1^2}{\overline{Q}_3^2}\right) 
+ \frac{Q_2}{4Q_1^2 \overline{Q}_3^3} \log \left( \frac{Q_1^2}{\overline{Q}_3^2}\right) \log \left( \frac{Q_2^2}{\overline{Q}_3^2}\right) \nonumber \\
&- \frac{Q_2}{4Q_1^2 \overline{Q}_3^3} \log ^2 \left( \frac{Q_1^2}{\overline{Q}_3^2}\right) 
+ \frac{Q_2}{Q_1^2 \overline{Q}_3^3} \mathcal{P}_3\left(1 + \frac{Q_2}{Q_1}\right) 
-  \frac{1}{2Q_1^2 \overline{Q}_3^2} \log \left( \frac{Q_1^2}{\overline{Q}_3^2}\right)\nonumber \\
&- \frac{1}{4 Q_1^2 \overline{Q}_3^2} \log \left( \frac{Q_1^2}{\overline{Q}_3^2}\right) \log \left( \frac{Q_2^2}{\overline{Q}_3^2}\right) 
+ \frac{5}{12Q_1^2 \overline{Q}_3^2} \log ^2 \left( \frac{Q_1^2}{\overline{Q}_3^2}\right)
- \frac{3}{2Q_1^2 \overline{Q}_3^2} \mathcal{P}_3\left(1 + \frac{Q_2}{Q_1}\right) \nonumber\\
&- \frac{1}{6Q_1^2 Q_2 \overline{Q}_3} \log ^2 \left( \frac{Q_1^2}{\overline{Q}_3^2}\right) \Bigg] + \mathcal{O}(\delta ^2)\, , 
\end{align}
\begin{align}
F_{123}(2) =&- 6 \zeta_3
- \frac{1}{2} \log \left( \frac{Q_1^2}{\overline{Q}_3^2}\right) \log ^2 \left( \frac{Q_2^2}{\overline{Q}_3^2}\right) 
+\frac{3}{2}\log ^2 \left( \frac{Q_1^2}{\overline{Q}_3^2}\right) \log \left( \frac{Q_2^2}{\overline{Q}_3^2}\right)
- \log ^3 \left( \frac{Q_1^2}{\overline{Q}_3^2}\right) \nonumber \\
&+ 6 \mathcal{P}_3\left(1 + \frac{Q_1}{Q_2} \right)
+  \delta \Bigg[  -  \frac{3Q_1}{Q_2 \overline{Q}_3^2} \log \left( \frac{Q_2^2}{\overline{Q}_3^2}\right)
+ \frac{Q_1}{Q_2 \overline{Q}_3^2} \log ^2 \left( \frac{Q_2^2}{\overline{Q}_3^2}\right) \nonumber \\
&+  \frac{3Q_1}{Q_2 \overline{Q}_3^2} \log \left( \frac{Q_1^2}{\overline{Q}_3^2}\right) 
-  \frac{5Q_1}{2Q_2 \overline{Q}_3^2} \log \left( \frac{Q_1^2}{\overline{Q}_3^2}\right) \log \left( \frac{Q_2^2}{\overline{Q}_3^2}\right) 
+\frac{3Q_1}{2Q_2 \overline{Q}_3^2} \log ^2 \left( \frac{Q_1^2}{\overline{Q}_3^2}\right) \nonumber \\
&+  \frac{3}{Q_2 \overline{Q}_3} \log \left( \frac{Q_2^2}{\overline{Q}_3^2}\right)
- \frac{1}{Q_2 \overline{Q}_3} \log ^2 \left( \frac{Q_2^2}{\overline{Q}_3^2}\right) 
+ \frac{5}{2Q_2 \overline{Q}_3} \log \left( \frac{Q_1^2}{\overline{Q}_3^2}\right) \log \left( \frac{Q_2^2}{\overline{Q}_3^2}\right)\nonumber \\
&- \frac{3}{2Q_2 \overline{Q}_3} \log ^2 \left( \frac{Q_1^2}{\overline{Q}_3^2}\right) \Bigg] + \mathcal{O}(\delta ^2)\, , 
\end{align}
\begin{align}
W_{123}(0) =&- \frac{3}{Q_2 \overline{Q}_3} \mathcal{P}_3\left(1 + \frac{Q_1}{ Q_2}\right) \nonumber \\
&+  \delta \Bigg[  \frac{Q_1}{2Q_2^2 \overline{Q}_3^3} \log \left( \frac{Q_2^2}{\overline{Q}_3^2}\right) 
-  \frac{Q_1}{4Q_2^2 \overline{Q}_3^3} \log ^2 \left( \frac{Q_2^2}{\overline{Q}_3^2}\right) 
- \frac{Q_1}{2Q_2^2 \overline{Q}_3^3} \log \left( \frac{Q_1^2}{\overline{Q}_3^2}\right)\nonumber \\
&+  \frac{Q_1}{4Q_2^2 \overline{Q}_3^3} \log \left( \frac{Q_1^2}{\overline{Q}_3^2}\right) \log \left( \frac{Q_2^2}{\overline{Q}_3^2}\right)
+ \frac{Q_1}{Q_2^2 \overline{Q}_3^3} \mathcal{P}_3\left(1 + \frac{Q_1}{Q_2}\right) 
- \frac{1}{2Q_2^2 \overline{Q}_3^2} \log \left( \frac{Q_2^2}{\overline{Q}_3^2}\right) \nonumber \\
&+ \frac{5}{12Q_2^2 \overline{Q}_3^2} \log ^2 \left( \frac{Q_2^2}{\overline{Q}_3^2}\right)
-\frac{1}{4Q_2^2 \overline{Q}_3^2} \log \left( \frac{Q_1^2}{\overline{Q}_3^2}\right) \log \left( \frac{Q_2^2}{\overline{Q}_3^2}\right)
- \frac{3}{2Q_2^2 \overline{Q}_3^2} \mathcal{P}_3\left(1 + \frac{Q_1}{Q_2} \right) \nonumber\\
&- \frac{1}{6Q_1 Q_2^2 \overline{Q}_3} \log ^2 \left( \frac{Q_2^2}{\overline{Q}_3^2}\right) \Bigg] \, . 
\end{align}
To obtain these one has to expand the relevant $\mathcal{P}_{i}$ functions around a general $z$. Note that one obtains e.g.~$\mathcal{P}_{3}(1+Q_2/Q1)$, which, from the definition of the function in~(\ref{eq:Pidefs}) gives rise to $\log \left( -Q_{2}/Q_{1}\right) $. This lies on the branch-cut of the (poly-)logarithm. However, the $\mathcal{P}_{i}$ are well-behaved, single-valued functions without branch-cuts and one can safely neglect these issues. 

When the $\hat{\Pi}$ are expanded, the negative powers of $\delta$ cancel. The expressions are not displayed here due to their length, but they can be found in the supplementary file {\tt resultsgluon.txt}. Equivalent expressions are provided for the quark loop in {\tt resultsquark.txt}.

\subsubsection{Corner regions}
In the corner regions the situation is different. There one has two small parameters instead of one
\begin{align}
\textrm{Corner Region }C_{i}: \; \; \; Q_{i}\ll Q_{j},Q_k \textrm{ and } \delta \equiv Q_{j}-Q_{k}\ll \overline{Q}_i\equiv Q_j+Q_k \, .
\end{align}
Below, we list the expansions in the region $C_3$ \footnote{In the supplementary file {\tt cornerexpansions.txt}, we provide the full expansions needed for all corner regions.}. The expansions of the master integrals are given by
\begin{equation}
C_{123}(0) =- \frac{8}{\overline{Q}_3^2} + \frac{8}{\overline{Q}_3^2} \log  \left(2\,\frac{Q_3}{\overline{Q}_3}\right)+ Q_3^2    \Bigg[  -  \frac{40}{9\overline{Q}_3^4} +  \frac{16}{3\overline{Q}_3^4} \log  \left( 2\,\frac{Q_3}{\overline{Q}_3}\right)\Bigg] +\mathcal{O}\left(Q_3^4,\delta^2,\delta Q_3^2\right)\,,
\end{equation}
\begin{equation}
W_{312}(0) =- \frac{24}{\overline{Q}_3^2} \zeta _3
+ Q_3^2    \Bigg[\frac{88}{3\overline{Q}_3^4} - \frac{16}{\overline{Q}_3^4} \zeta _3 - \frac{16}{\overline{Q}_3^4}  \log  \left(2\,\frac{Q_3}{\overline{Q}_3}\right) \Bigg]+\mathcal{O}\left(Q_3^4,\delta^2,\delta Q_3^2\right)\,,
\end{equation}
\begin{equation}
F_{312}(2) = 6 \zeta _3 - 8 \log ^3 \left(2\, \frac{Q_3}{\overline{Q}_3}\right)
+ Q_3^2    \Bigg[- \frac{36}{\overline{Q}_3^2} + \frac{24}{\overline{Q}_3^2} \log  \left(2\frac{Q_3}{\overline{Q}_3}\right) \Bigg]+\mathcal{O}\left(Q_3^4,\delta^2,\delta Q_3^2\right)
\end{equation}
\begin{align}
W_{213}(0) =&- \frac{24}{\overline{Q}_3^2} + \frac{24}{\overline{Q}_3^2}  \log  \left(2\, \frac{Q_3}{\overline{Q}_3}
\right) -  \frac{8}{\overline{Q}_3^2}  \log ^2 \left(2\, \frac{Q_3}{\overline{Q}_3}\right) \nonumber\\
&+ Q_3^2    \Bigg[ -  \frac{130}{27\overline{Q}_3^4} +  \frac{76}{9\overline{Q}_3^4} \log \left( 2\,\frac{Q_3}{\overline{Q}_3}\right) -  \frac{40}{9\overline{Q}_3^4}  \log ^2 \left(2\,\frac{Q_3}{\overline{Q}_3}\right)\Bigg] \nonumber \\
&+ \delta    \Bigg[ \frac{18}{\overline{Q}_3^3} -  \frac{20}{\overline{Q}_3^3} \log  \left(2\, \frac{Q_3}{\overline{Q}_3}
\right) +  \frac{8}{\overline{Q}_3^3}  \log ^2 \left(2\, \frac{Q_3}{\overline{Q}_3}\right) \Bigg]
+\mathcal{O}\left(Q_3^4,\delta^2,\delta Q_3^2\right)\,, 
\end{align}
\begin{align}
F_{213}(2) =&- 6 \zeta _3
+ Q_3^2    \Bigg[\frac{18}{\overline{Q}_3^2} - \frac{12}{\overline{Q}_3^2}  \log  \left(2\,\frac{Q_3}{\overline{Q}_3}
\right) \Bigg] \nonumber \\
&+ \delta    \Bigg[\frac{24}{\overline{Q}_3} - \frac{24}{\overline{Q}_3} \log  \left(2\,\frac{Q_3}{\overline{Q}_3}\right)
+  \frac{16}{\overline{Q}_3} \log ^2 \left(2\,\frac{Q_3}{\overline{Q}_3}\right) \Bigg]
+\mathcal{O}\left(Q_3^4,\delta^2,\delta Q_3^2\right)\,, 
\end{align}
\begin{align}
W_{123}(0) =&-  \frac{24}{\overline{Q}_3^2} + \frac{24}{\overline{Q}_3^2}  \log  \left(2\,\frac{Q_3}{\overline{Q}_3}
\right) -  \frac{8}{\overline{Q}_3^2}  \log  ^2 \left(2\,\frac{Q_3}{\overline{Q}_3}\right) \nonumber\\
&+ Q_3^2    \Bigg[- \frac{130}{27\overline{Q}_3^4} +  \frac{76}{9\overline{Q}_3^4}  \log  \left(2\,\frac{Q_3}{\overline{Q}_3}
\right) - \frac{40}{9\overline{Q}_3^4}  \log ^2  \left(2\,\frac{Q_3}{\overline{Q}_3}\right)\Bigg] \nonumber \\
&+ \delta    \Bigg[- \frac{18}{\overline{Q}_3^3} +  \frac{20}{\overline{Q}_3^3}  \log  \left(2\,\frac{Q_3}{\overline{Q}_3}
\right) - \frac{8}{\overline{Q}_3^3}  \log ^2  \left(2\,\frac{Q_3}{\overline{Q}_3}
\right) \Bigg]
+\mathcal{O}\left(Q_3^4,\delta^2,\delta Q_3^2\right)\,, 
\end{align}
\begin{align}
F_{123}(2) =&- 6 \zeta _3
+ Q_3^2    \Bigg[\frac{18}{\overline{Q}_3^2} -  \frac{12}{\overline{Q}_3^2} \log  \left(2\,\frac{Q_3}{\overline{Q}_3}
\right) \Bigg]\,, \nonumber \\
&+ \delta    \Bigg[ -  \frac{24}{\overline{Q}_3} + \frac{24}{\overline{Q}_3}  \log  \left(2\,\frac{Q_3}{\overline{Q}_3}
\right) - \frac{16}{\overline{Q}_3} \log ^2 \left(2\frac{Q_3}{\overline{Q}_3}
\right) \Bigg]
+\mathcal{O}\left(Q_3^4,\delta^2,\delta Q_3^2\right)\,. 
\end{align}
In the corner regions one has to expand the $\mathcal{P}_{i}(z)$ for three different $z$, namely $z=0,\, 1,\, \infty$. However, from the relations in~(\ref{eq:Pirels}) one can relate $\mathcal{P}_{i}(z)$ to $\mathcal{P}_{i}(1/z)$, so only $z=0,1$ are needed in practice. 

The $\hat{\Pi}$ in region $C_{3}$ are given by 

\begin{align}
\frac{\hat{\Pi}_1}{c_s}=&
\frac{1}{Q_3^2}   \Bigg[ \frac{192}{\overline{Q}_3^2} \Bigg]
+ \frac{3568}{15\overline{Q}_3^4} -\frac{2048}{5\overline{Q}_3^4} \zeta_3 + \frac{3776}{9\overline{Q}_3^4} \log  \left( 2\,\frac{Q_3}{\overline{Q}_3}\right) - 
 \frac{256}{\overline{Q}_3^4} \log ^2 \left( 2\,\frac{Q_3}{\overline{Q}_3}\right)\nonumber \\
&+ Q_3^2  \Bigg[  \frac{39375808}{39375\overline{Q}_3^6} - \frac{36864}{35\overline{Q}_3^6} \zeta_3 +  \frac{528896}{1125\overline{Q}_3^6} \log  \left( 2\,\frac{Q_3}{\overline{Q}_3}\right) -\frac{12288}{25\overline{Q}_3^6} \log ^2 \left( 2\,\frac{Q_3}{\overline{Q}_3}\right) \Bigg]\nonumber \\
&+  \frac{\delta^2}{Q_3^2}   \Bigg[  \frac{1280}{9\overline{Q}_3^4} - \frac{2560}{9\overline{Q}_3^4} \log  \left( 2\,\frac{Q_3}{\overline{Q}_3}\right) \Bigg]
+\mathcal{O}(\delta ^2,\delta ^4 Q_3^{-2})\,,
\end{align}
\begin{align}
\frac{\hat{\Pi}_4}{c_s} =&
- \frac{12544}{45\overline{Q}_3^4} + \frac{2048}{5\overline{Q}_3^4} \zeta_3 + \frac{2048}{9\overline{Q}_3^4} \log  \left( 2\,\frac{Q_3}{\overline{Q}_3}\right) \nonumber\\
&+ Q_3^2   \Bigg[  - \frac{13533568}{13125\overline{Q}_3^6} + \frac{49152}{35\overline{Q}_3^6} \zeta_3 + 
  \frac{1935872}{1125\overline{Q}_3^6} \log  \left( 2\,\frac{Q_3}{\overline{Q}_3}\right) -  \frac{14336}{25\overline{Q}_3^6} \log ^2 \left( 2\,\frac{Q_3}{\overline{Q}_3}\right) \Bigg] \nonumber \\
&+\mathcal{O}(\delta ^2,Q_3^2,\delta ^2 Q_3^2)\,,
\end{align}
\begin{align}
\frac{\hat{\Pi}_7}{c_s} =&- \frac{1024}{5\overline{Q}_3^6} + \frac{8192}{5\overline{Q}_3^6} \zeta_3 + \frac{8192}{9\overline{Q}_3^6}\log  \left( 2\,\frac{Q_3}{\overline{Q}_3}\right) -  \frac{\delta}{Q_3^2}   \Bigg[   \frac{1024}{9\overline{Q}_3^5} \Bigg] \nonumber \\
&+ \delta   \Bigg[  - \frac{78295808}{13125\overline{Q}_3^7} + \frac{196608}{35\overline{Q}_3^7} \zeta_3 -
  \frac{17408}{1125\overline{Q}_3^7} \log  \left( 2\,\frac{Q_3}{\overline{Q}_3}\right) + \frac{69632}{75\overline{Q}_3^7} \log ^2 \left( 2\,\frac{Q_3}{\overline{Q}_3}\right) \Bigg]\nonumber \\
&+Q_3^2   \Bigg[  - \frac{128872192}{39375 \overline{Q}_3^8} + \frac{98304}{35\overline{Q}_3^8} \zeta_3 +  \frac{13229056}{1125\overline{Q}_3^8} \log  \left( 2\,\frac{Q_3}{\overline{Q}_3}\right) - \frac{323584}{75\overline{Q}_3^8} \log ^2 \left( 2\,\frac{Q_3}{\overline{Q}_3}\right) \Bigg] \nonumber \\
&+\mathcal{O}(\delta ^2,\delta Q_3^2,\delta ^3 Q_3^{-2})\,,
\end{align}
\begin{align}
\frac{\hat{\Pi}_{17}}{c_s} =&
\frac{1}{Q_3^2}   \Bigg[  - \frac{512 }{\overline{Q}_3^4} +  \frac{1024}{9\overline{Q}_3^4} \log  \left( 2\,\frac{Q_3}{\overline{Q}_3}\right) \Bigg] \nonumber \\
&- \frac{112018624}{118125\overline{Q}_3^6} + \frac{45056}{35\overline{Q}_3^6} \zeta_3 - \frac{1769216}{1125\overline{Q}_3^6} 
 \log  \left( 2\,\frac{Q_3}{\overline{Q}_3}\right) + \frac{115712}{225\overline{Q}_3^6} \log ^2 \left( 2\,\frac{Q_3}{\overline{Q}_3}\right) \nonumber  \\
&+ Q_3^2   \Bigg[  - \frac{139984302592}{24310125\overline{Q}_3^8} + \frac{212992}{35\overline{Q}_3^8} \zeta_3 - 
 \frac{52059136}{25725\overline{Q}_3^8}\log  \left( 2\,\frac{Q_3}{\overline{Q}_3}\right) + \frac{1101824}{735\overline{Q}_3^8}\log ^2 \left( 2\,\frac{Q_3}{\overline{Q}_3}\right) \Bigg] \nonumber \\
&+  \frac{\delta^2}{Q_3^2}   \Bigg[ -\frac{336896}{1125 \overline{Q}_3^6} +  \frac{47104}{225\overline{Q}_3^6} \log  \left( 2\,\frac{Q_3}{\overline{Q}_3}\right) \Bigg]
+\mathcal{O}(\delta ^2,\delta ^4 Q_3^{-2})\,,
\end{align}
\begin{align}
\frac{\hat{\Pi}_{39}}{c_s} =&
 \frac{1}{Q_3^2}   \Bigg[  - \frac{1024}{9\overline{Q}_3^4} +\frac{1024}{9\overline{Q}_3^4} \log  \left( 2\,\frac{Q_3}{\overline{Q}_3}\right) \Bigg] \nonumber \\
&- \frac{53725888}{118125 \overline{Q}_3^6} - \frac{4096}{35\overline{Q}_3^6} \zeta_3 + \frac{725248}{1125\overline{Q}_3^6} 
 \log  \left( 2\,\frac{Q_3}{\overline{Q}_3}\right) - \frac{80896}{225\overline{Q}_3^6} \log ^2 \left( 2\,\frac{Q_3}{\overline{Q}_3}\right)\nonumber \\
&+ Q_3^2   \Bigg[ - \frac{75027492352}{121550625\overline{Q}_3^8} - \frac{16384}{35\overline{Q}_3^8} \zeta_3 - 
 \frac{356996096}{385875 \overline{Q}_3^8} \log  \left( 2\,\frac{Q_3}{\overline{Q}_3}\right) -  \frac{167936}{1225\overline{Q}_3^8} \log ^2 \left( 2\,\frac{Q_3}{\overline{Q}_3}\right) \Bigg] \nonumber \\
&+  \frac{\delta^2}{Q_3^2}   \Bigg[ \frac{11264}{1125\overline{Q}_3^6} + \frac{47104}{225\overline{Q}_3^6} \log  \left( 2\,\frac{Q_3}{\overline{Q}_3}\right)\Bigg]
+\mathcal{O}(\delta ^2,\delta ^4 Q_3^{-2})\,,
\end{align}
\begin{align}
\frac{\hat{\Pi}_{54}}{c_s} =&
 \frac{\delta}{Q_3^2}   \Bigg[  -  \frac{1408}{9\overline{Q}_3^5} - \frac{1024}{3\overline{Q}_3^5} \log  \left( 2\,\frac{Q_3}{\overline{Q}_3}\right) \Bigg] \nonumber \\
&+ \delta  \Bigg[  - \frac{1937408}{1575\overline{Q}_3^7} + \frac{24576}{35\overline{Q}_3^7} \zeta_3 - 
 \frac{687104}{225\overline{Q}_3^7} \log  \left( 2\,\frac{Q_3}{\overline{Q}_3}\right) + \frac{4096}{5\overline{Q}_3^7} \log ^2 \left( 2\,\frac{Q_3}{\overline{Q}_3}\right) \Bigg] \nonumber \\
&+\mathcal{O}(\delta Q_3^2,\delta ^3 Q_3^{-2},\delta ^3)\,,
\end{align}
with the overall factor $c_s=\frac{2\pi \alpha _s (N_c^2-1) e_q^4}{(16\pi ^2)^2}$.
While all these expressions are finite when the small parameter $\delta$ tends to zero, some of them diverge when $Q_3^2\to 0$. However, this divergence has no physical meaning, since that limit lies outside the region of validity of the OPE. The expansion of the $\hat{\Pi}$ in the other corner regions can be found in the supplementary file {\tt resultsgluon.txt}. Equivalent expressions are provided for the quark loop in {\tt resultsquark.txt}.

\subsection{Symmetric Point}\label{app:symmetricpoint}

In this section, we write the 
expressions for the $\hat{\Pi}$ at the symmetric point $Q_1=Q_2=Q_3=Q$.\footnote{The associated $\tilde{\Pi}$ are also available upon request.} For the quark loop these are
\begin{align}
&\frac{16 \pi ^2 \hat{\Pi}_1^\text{quark}}{N_c   e_q^4} = - \frac{32}{3\,Q^4} \, , \nonumber  \\
&\frac{16 \pi ^2\hat{\Pi}_4^\text{quark}}{N_c   e_q^4} =\frac{2}{Q^4}\left( - \frac{352}{27} + \frac{128}{81} \, \Delta^{(1)} \right) \, , \nonumber  \\
&\frac{16 \pi ^2\hat{\Pi}_7^\text{quark} }{N_c   e_q^4}=\frac{1}{Q^6}\left( - \frac{352}{27} +  \frac{128}{81} \,\Delta^{(1)} \right) \, , \nonumber  \\
&\frac{16 \pi ^2\hat{\Pi}_{17}^\text{quark} }{N_c   e_q^4}=\frac{1}{Q^6}\left( - \frac{384}{27} + \frac{192}{81} \, \Delta^{(1)}\right) \, , \nonumber  \\
&\frac{16 \pi ^2\hat{\Pi}_{39}^\text{quark}}{N_c   e_q^4} = \frac{1}{Q^6}\left( \frac{320}{27} -  \frac{64}{81} \, \Delta^{(1)}\right) \, , \nonumber  \\
&\frac{16 \pi ^2 \hat{\Pi}_{54}^\text{quark}}{N_c   e_q^4} = 0 \, ,
\end{align}
where
\begin{equation}
\Delta^{(n)}\equiv \psi ^{(n)}(1/3)- \psi ^{(n)}(2/3) \, .
\end{equation}
and $\psi ^{(n)}$ is the polygamma function of order $n$ defined by
\begin{equation}
    \psi ^{(n)}(z) \equiv \frac{d^{n+1}}{dz^{n+1}}\log \Gamma (z)\,.
\end{equation}
One then has
\begin{align}
\Delta^{(1)}&\approx 7.031721716 \, ,  \\
\Delta^{(3)}&\approx 456.8524809 \, .
\end{align}
These expressions agree with those given in Ref.~\cite{Colangelo:2019uex}. For the gluonic correction the result is
\begin{align}
&\frac{\hat{\Pi}_1^\text{gluon}}{c_s} = \frac{1}{Q^4}\left(\frac{640 \zeta _3}{3} + \frac{400}{9} \, \Delta^{(1)}  - \frac{32}{27}  \,  \Delta^{(3)} \right)  \, , \nonumber  \\
&\frac{\hat{\Pi}_4^\text{gluon}}{c_s} = \frac{1}{Q^4}\left(-  32 + \frac{1664 \zeta _3}{3} + \frac{1040}{9} \,\Delta^{(1)} - \frac{256}{81} \,  \Delta^{(3)} \right)  \, , \nonumber  \\
&\frac{\hat{\Pi}_7^\text{gluon}}{c_s} =\frac{1}{Q^6}\left(  \frac{512\zeta _3}{3} +\frac{320}{9}\, \Delta^{(1)} - \frac{80}{81}\,  \Delta^{(3)}\right)  \, , \nonumber  \\
&\frac{\hat{\Pi}_{17}^\text{gluon}}{c_s} =\frac{1}{Q^6}\left(  - 32 + 384 \zeta_3 +  \frac{272}{3}\,  \Delta^{(1)} - \frac{64}{27} \,  \Delta^{(3)} \right)\, , \nonumber  \\
&\frac{\hat{\Pi}_{39}^\text{gluon}}{c_s} = \frac{1}{Q^6}\left( - \frac{32}{3} -  128 \zeta _3 - \frac{272}{9}\,  \Delta^{(1)}  + \frac{64}{81} \, \Delta^{(3)} \right)\, , \nonumber  \\
&\frac{\hat{\Pi}_{54}^\text{gluon}}{c_s} = 0 \, ,
\end{align}
where $c_s=\frac{2\pi \alpha _s (N_c^2-1) e_q^4}{(16\pi ^2)^2}$.


\providecommand{\href}[2]{#2}\begingroup\raggedright\endgroup

\end{document}